\newcommand{\exi}{\ensuremath{\mathbf{x}_i}}
\newcommand{\yij}{\ensuremath{\mathbf{y}_{ij}}}
\newcommand{\sij}{\ensuremath{\sigma_{ij}}}
\newcommand{\odds}{\ensuremath{\mathcal{O}}}

\documentclass[12pt]{article}
\usepackage{aasms4}
\begin{document}

\title{Evidence Against an Association Between Gamma-Ray Bursts and
Type I Supernovae}

\author{C. Graziani, D. Q. Lamb and G. H. Marion}

\affil{Department of Astronomy and Astrophysics, University
of Chicago, \\ 5640 South Ellis Avenue, Chicago, IL 60637}

\begin{abstract}
We present a rigorous method, based on Bayesian inference, for
calculating the odds favoring the hypothesis that any particular class
of astronomical transients produce gamma-ray bursts over the hypothesis
that they do not.  We then apply this method to a sample of 83 Type Ia 
supernovae and a sample of 20 Type Ib-Ic supernovae.  We find
overwhelming odds against the hypothesis that all Type Ia supernovae
produce gamma-ray bursts, whether at low redshift ($10^{9}:1$)  or
high-redshift ($10^{12}:1$), and very large odds ($6000:1$) against the
hypothesis that all Type Ib, Ib/c, and Ic supernovae produce observable
gamma-ray bursts.  We find large odds ($34:1$) against the hypothesis
that a fraction of Type Ia supernovae produce observable gamma-ray
bursts, and moderate odds ($6:1$) against the hypothesis that a
fraction of Type Ib-Ic supernovae produce observable bursts.  We have
also re-analyzed both a corrected version of the \markcite{wang98}Wang \&
Wheeler sample of Type Ib-Ic SNe and our larger sample of 20 Type Ib-Ic
SNe, using a generalization of their frequentist method.  We find no
significant evidence in either case of a correlation between Type Ib-Ic
SNe and GRBs, consistent with the very strong evidence against such a
correlation that we find from our Bayesian analysis.
\end{abstract}


\section{Introduction}

The discovery that the sky distribution of faint gamma-ray bursts
(GRBs) is isotropic, coupled with the confirmation of a roll-over in
the cumulative brightness distribution of the bursts, suggested that
the bursts lie at cosmological distances (\cite{meegan92}).  About one
year ago, the rapid dissemination of arcminute-sized GRB error circles
provided by the Wide-Field Camera (WFC) on BeppoSAX (\cite{costa97a})
led to the discovery of fading X-ray (\cite{costa97b}) and optical
(\cite{groot97}) counterparts to the bursts.  The subsequent
measurement of absorption lines at $z = 0.835$ in the spectra of the
optical afterglow of GRB 970508 (\cite{metzger97}) firmly established
the extra-galactic nature of this burst, and presumably, of most or all
GRBs.  Redshifts  are now known for the apparent host galaxies of two
other bursts:  $z = 0.965$ in the case of GRB 980703
(\cite{djorgovski98}) and $z = 3.42$ in the case of GRB 971214
(\cite{kulkarni98a}).  

GRB980425 has complicated this simple ``cosmological'' picture of
GRBs.  Following the detection of this burst by the BeppoSAX Gamma-Ray
Burst Monitor and WFC (\cite{soffitta98}; see also \cite{kippen98a}),
X-ray follow-up observations were made using the BeppoSAX Narrow Field
Instrument (NFI) \markcite{pian98a,pian98b,piro98}(Pian et al. 1998a,b;
Piro et al. 1998).  The initial observation revealed a faint X-ray
source (detected at the $5\sigma$ level) that was not seen in several
subsequent observations.  Optical follow-up observations led to the
discovery of a supernova, SN 1998bw, within the 8 arcminute radius of
the BepppoSAX WFC error circle for the GRB but not coincident with the
fading X-ray source (\cite{galama98}).  No other optically variable
object was detected within the BeppoSAX WFC error circle for the
burst.  SN 1998bw was subsequently found to be of Type Ic
(\cite{galama98}) and very bright in the radio (\cite{kulkarni98b}). 
The supernova is coincident with a galaxy (presumably the host galaxy)
that lies at $z = 0.008$ (\cite{tinney98}).  

An association between GRB 980425 and SN 1998bw is an intriguing
possibility, made more so by the recent heightened interest in
``collapsar'' or ``hypernova'' models of GRBs
(\markcite{woosley93,woosley98}Woosley 1993; Woosley, Eastman \&
Schmidt 1998; \cite{paczynski98}; \cite{hoflich98}). The principal
argument in favor of an association between GRB 980425 and SN 1998bw is
the positional and temporal coincidence between the two events.  Given
a supernova rate of $\sim2$ per $L_\star$ galaxy per century
(\cite{strom95}), a density of $L_\star$ galaxies of 0.01 Mpc$^{-3}$,
and that approximately 2/7 of these are SNe of Types Ib, Ib/c and Ic
(\cite{woosley86}, \cite{strom95}), the chance probability of such a
spatial-temporal coincidence for a Type Ib-Ic SN with $z\le 0.008$ is
$\sim 10^{-7}$.  But would an association between GRB980425 and
SN 1998bw be acceptable SN 1998bw were at $z=1$ rather than at
$z=0.008$?   Almost certainly.  If so, the appropriate value for the
chance probability of the positional and temporal coincidence becomes
$\sim 0.1$\%.  This illustrates how difficult it is to evaluate {\it a
posteriori} statistical arguments.

And there are specific reasons to be cautious in this case.  Assuming a
power-law decay with time, and connecting the 2-10 keV X-ray flux
detected by the BeppoSAX WFC during and immediately following the
burst, and the 2-10 keV X-ray flux of the fading X-ray source detected
10 hours later by the BeppoSAX NFI, yields a power-law index of $\sim
1.2$ (\cite{pian98b}), which is similar to the power-law indices of the
X-ray afterglows of other BeppoSAX bursts.  Thus GRB 980425 is more
plausibly associated with this fading X-ray source than with SN 1998bw.

Also, if the association between GRB 980425 and SN 1998bw were true,
the luminosity of this burst would be $\sim 10^{46}$ erg s$^{-1}$ and
its energy  would be $\sim 10^{47}$ erg.  Each would therefore be five
orders of magnitude less than that of other bursts, and the behavior of
the X-ray and optical afterglow would be very different from those of
the other BeppoSAX bursts, yet the burst itself is indistinguishable
from other BeppoSAX and BATSE GRBs with respect to duration, time
history, spectral shape, peak flux, and and fluence (\cite{galama98}).

In view of the difficulty in assessing the significance of any
association between SNe and GRBs on the basis of this single event, the
safest procedure is to regard the association as a hypothesis that is
to be tested by searching for correlations between SNe and GRB in
catalogs of SNe and GRBs, excluding SN 1998bw and GRB980425.  Wang \&
Wheeler (1998) have performed such a study, and find evidence for a
significant (at the $10^{-5}$ level) correlation between Type Ib-Ic SNe 
and GRBs detected by BATSE.

While the results of \markcite{wang98}Wang \& Wheeler (1998) seem 
promising, their study suffers from several deficiencies.  The number
(six) of Type Ib-Ic SNe in their sample is small, and one of these
events is mis-classified [SN 1992ad is a Type II SN
(\cite{mcnaught92,filippenko92}), not a Type Ic SN], which eliminates
one of their SN-GRB associations.  The range of possible explosion
dates that we derive for another event (SN 1997X) is  much smaller than
the range they allow, which rules out another of their  associations. 
Furthermore, two other SN--GRB associations are ruled out by
Interplanetary Network positions (\cite{hurley98,kippen98a}). 
Moreover, \markcite{wang98}Wang \& Wheeler's methodology is somewhat
arbitrary, in the sense that they increase the size of the BATSE GRB
positional error circles by a large, arbitrary factor.  Finally,
\markcite{wang98}Wang \& Wheeler's methodology makes no provision for
the fact that the BATSE temporal exposure is less than unity.  In fact,
their result (six of six ``Type Ib-Ic'' SNe correlated with GRBs) is
unlikely, even if the proposed association between Type Ib-Ic SNe and
GRBs were real, since BATSE has on average a probability of 0.48 of
detecting any given GRB because of Earth blocking and other effects
(\cite{hakkila98}).

Here we carry out an analysis that overcomes these deficiencies.  We
correct the ``Type Ib-Ic'' SN sample of \markcite{wang98}Wang \&
Wheeler (1998) and supplement it with 15 additional Type Ib-Ic SNe, so
that we can study a larger sample.   Further, we develop an alternative
method, based on Bayesian inference and therefore using the likelihood
function, that incorporates information about the BATSE position errors
in a non-arbitrary way and that is free of the ambiguities of {\it a
posteriori} statistics.  The method also accounts the fact that the
BATSE temporal exposure is less than unity.

Applying this method to a sample of 83 Type Ia SNe and a sample of 20
Type Ib-Ic SNe, we find overwhelming odds against the hypothesis that
all Type Ia SNe produce observable gamma-ray bursts, irrespective of
whether the SNe are at low- or high-redshift, and very large odds
against the hypothesis that all Type Ib, Ib/c, and Ic SNe produce
observable gamma-ray bursts.  We find large odds against the hypothesis
that a fraction of Type Ia supernovae produce observable gamma-ray
bursts, and moderate odds against the hypothesis that a fraction of
Type Ib, Ib/c, and Ic supernovae produce observable bursts.  

We have also re-analyzed both a corrected version of the
\markcite{wang98}Wang \& Wheeler sample of Type Ib-Ic SNe and our
larger sample of 20 Type Ib-Ic SNe, using a generalization of their
frequentist method.  We find no significant evidence in either case of
a correlation between Type Ib-Ic SNe and GRBs, consistent with the very
strong evidence against such a correlation that we find from our
Bayesian analysis. 

The plan of this paper is as follows.  In \S II we present a rigorous
method, based on Bayesian inference, for calculating the odds favoring
the hypothesis that any particular class of astronomical transients 
produces GRBs over the hypothesis that they do not.  In \S III we apply
this method to various subclasses of Type I SNe.  In \S IV we discuss
our results, and compare them with other work.  We present our
conclusions in \S V.

\section{Statistical Methodology}

\subsection{Bayesian Odds}

We denote the data by $D=\{D_i | i=1,\ldots N_{\mathrm{SN}}\}$, where
$N_{\mathrm{SN}}$ is the number of observed SNe.  For the $i$th
observed SN, the data consists of the SN position $\exi$
(a unit vector), the earliest time $t_i$ at which the SN explosion could
have occurred, the duration $\tau_i$ of the period of time during which
the SN explosion could have occurred, the number $N_i$ of GRBs that
occurred during the time interval $[t_i,t_i+\tau_i]$, and the list
$(\yij,\sij), j=1,\ldots,N_i$ of BATSE positions and error parameters
for those bursts.  Thus, $D_i=\{\exi,t_i,\tau_i,N_i,\{\,(\yij,\sij)\,|
\,j=1,\ldots,N_i\}\,\}$.

Note that the $\sij$ are Fisher distribution parameters, \emph{not} the
BATSE-style 68\% error-circle radii.  They enter the odds calculation
through the assumption that an observed burst position $\mathbf{y}$ is
distributed around its true position $\mathbf{x}$ according to the
Fisher distribution
\begin{mathletters}
\begin{eqnarray}
P(\mathbf{y}|\mathbf{x},\sigma)&=&
\kappa\exp\left[(\mathbf{y}\cdot\mathbf{x}-1)/\sigma^2\right]\\
\kappa&\equiv&\left[2\pi\sigma^2\left(1-e^{-2/\sigma^2}\right)\right]^{-1}
\end{eqnarray}
\end{mathletters}
(see, for example \cite{mardia}, p. 228).  The $\sigma$ are related to the
68\% total errors (including correction for systematic error) by
the linear relation $\sigma_{\mathrm{tot}}^{68\%}=1.52\sigma$.

We compare two hypotheses:

\noindent $H_1$: The association between SNe and GRBs is real.  If a SN
is observed, there is a chance $\epsilon$ that BATSE sees the
associated GRB, where $\epsilon$ is the average BATSE temporal
exposure.  While $\epsilon$ varies with Declination, the variation is
modest and we neglect it.  The probability density for the time of
occurrence of the $i$th supernova is assumed uniform in the interval
$[t_i,t_i+\tau_i]$, so that all GRBs that occur in that interval have
an equal prior probability of being associated with the SN.

\noindent $H_2$: There is no association between SNe and GRBs.

We wish to calculate the odds favoring $H_1$ over $H_2$, given the data.
That is, we want
\begin{eqnarray}
\odds&\equiv&\frac{P(H_1|D,I)}{P(H_2|D,I)}\nonumber\\
&=&\frac{P(D|H_1,I)\,P(H_1|I)}{P(D|H_2,I)\,P(H_2|I)}\nonumber\\
&=&\frac{P(D|H_1,I)}{P(D|H_2,I)}\nonumber\\
&=&\prod_{i=1}^{N_{\mathrm{sn}}}\frac{P(D_i|H_1,I)}{P(D_i|H_2,I)},
\label{odds1}
\end{eqnarray}
where we have set the prior probabilities $P(H_1|I)=P(H_2|I)=1/2$, and we
have assumed the statistical independence of all the $D_i$.  The symbol
$I$ is shorthand for all the available prior information.

From equation (\ref{odds1}), it is apparent that $\odds$ is equal to
the likelihood ratio.  A simplification that occurs here is that our
hypotheses $H_1$ and $H_2$ are simple --- they are not parametrized
families of models.  As a consequence, the likelihoods $P(D|H_i,I)$ are
not ``global'' likelihoods, averaged over parameter space weighted by a
prior density, as is common in odds ratio calculations (\cite{ll92},
\cite{graziani92}).  Rather, they are genuine likelihoods, the
computation of which requires no prior probability density over
parameter space.

\subsection{Simple Model}

We now compute the likelihoods for a model in which all Type Ib-Ic SNe
produce GRBs.  Under the no-association hypothesis $H_2$, we have
\begin{eqnarray}
P(D_i|H_2,I)&=&P(\exi,t_i,\tau_i|H_2,I)\times
P(N_i,\{\yij,\sij\}|\tau_i,H_2,I)\nonumber\\
&=&f(\exi,t_i,\tau_i)\times \frac{e^{-R\tau_i}(R\tau_i)^{N_i}}{N_i!}
\times\left[\prod_{j=1}^{N_i}g(\sij)\right]
\times \left(\frac{1}{4\pi}\right)^{N_i},
\label{ph2}
\end{eqnarray}
where $f(\exi,t_i,\tau_i)\,d^2\exi\,dt_i\,d\tau_i$ is the differential
rate for observing such SNe, $g(\sigma)\,d\sigma$ is the differential
rate for observing a BATSE GRB positional error $\sigma$,
$(4\pi)^{-N_i}\,d^2\mathbf{y}_{i1}\ldots d^2\mathbf{y}_{iN_i}$ is the
differential probability of $N_i$ isotropic GRB positions, and $R$ is
the time rate at which BATSE observes GRBs.  It is unnecessary to
specify $f$ and $g$ in greater detail, since they are the same under
$H_1$ as under $H_2$, so that they cancel in the odds.

Under the association hypothesis $H_1$, we must take into account the
possibility that the GRB associated with the $i$th SN may not have been
detected by BATSE as a consequence of incomplete temporal exposure. We
denote by $E$ the proposition that BATSE was exposed to the direction
of the SN when it occurred, and by $\bar{E}$ the negation of $E$. 
Then,
\begin{eqnarray}
P(D_i|H_1,I)&=&P(D_i,E|H_1,I)+P(D_i,\bar{E}|H_1,I)\nonumber\\
&=&P(\bar{E}|H_1,I)P(D_i|\bar{E},H_1,I)+P(E|H_1,I)P(D_i|E,H_1,I)\nonumber\\
&=&(1-\epsilon)P(D_i|\bar{E},H_1,I)+\epsilon P(D_i|E,H_1,I).
\label{ph1}
\end{eqnarray}

Now, if BATSE was not exposed to the SN, then the $N_i$ observed GRBs
are purely coincidental, and the probability for observing them is the
same as it would be if $H_2$ held instead of $H_1$:
\begin{equation}
P(D_i|\bar{E},H_1,I)=P(D_i|H_2,I).
\label{h1h2}
\end{equation}

The second term in equation (\ref{ph1}) is
\begin{eqnarray}
P(D_i|E,H_1,I)&=&
P(\exi,t_i,\tau_i|H_1,I)\times P(N_i|\tau_i,E,H_1,I)\times
P(\{\sij\}|N_i,E,H_1,I)\nonumber\\
&&\times\,
P(\{\yij\}|\{\sij\},N_i,\exi,E,H_1,I)\\
&=&f(\exi,t_i,\tau_i)\times \frac{e^{-R\tau_i}(R\tau_i)^{N_i-1}}{(N_i-1)!}
\times \left(\prod_{j=1}^{N_i}g(\sij)\right)\nonumber\\
&&\times\, P(\{\yij\}|\{\sij\},N_i,\exi,E,H_1,I).
\label{expo}
\end{eqnarray}

Denoting by $A_{ij}$ the proposition that the $j$th GRB is
associated with the $i$th SN, we have
\begin{eqnarray}
P(\{\yij\}|\{\sij\},N_i,\exi,\tau_i,E,H_1,I)&=&
\sum_{j=1}^{N_i}P(A_{ij},\{\yij\}|\{\sij\},N_i,\exi,E,H_1,I)\nonumber\\
&=&\sum_{j=1}^{N_i}P(A_{ij}|N_i,E,H_1,I)\nonumber\\
&&\times\, P(\{\yij\}|A_{ij},\{\sij\},N_i,\exi,E,H_1,I)\nonumber\\
&=&\sum_{j=1}^{N_i}\frac{1}{N_i}
\left(\frac{1}{4\pi}\right)^{N_i-1}\times
\frac{\exp\left[(\yij\cdot\exi-1)/\sij^2\right]}
{2\pi\sij^2\left(1-e^{-2/\sij^2}\right)},
\label{alik}
\end{eqnarray}
where we have used the assumed equality of the prior probabilities
$P(A_{ij}|N_i,E,H_1,I)=1/N_i$, as well as the Fisher distribution for
the position of the GRB associated with the SN.

Combining equations (\ref{expo}) and (\ref{alik}), we obtain
\begin{eqnarray}
P(D_i|E,H_1,I)&=&
f(\exi,t_i,\tau_i)\times \frac{e^{-R\tau_i}(R\tau_i)^{N_i-1}}{(N_i-1)!}
\times \left(\prod_{j=1}^{N_i}g(\sij)\right)
\times \frac{1}{N_i}\left(\frac{1}{4\pi}\right)^{N_i-1}\nonumber\\
&&\times\, \sum_{j=1}^{N_i}
\frac{\exp\left[(\yij\cdot\exi-1)/\sij^2\right]}
{2\pi\sij^2\left(1-e^{-2/\sij^2}\right)} \nonumber\\
&=&P(D_i|H_2,I)\times\frac{1}{R\tau_i}
\times \sum_{j=1}^{N_i}\frac{\exp\left[(\yij\cdot\exi-1)/\sij^2\right]}
{\frac{1}{2}\sij^2\left(1-e^{-2/\sij^2}\right)}.
\label{expo2}
\end{eqnarray}

Finally, inserting equations (\ref{expo2}) and (\ref{h1h2}) into
equation (\ref{ph1}), and combining the result with equation
(\ref{odds1}), we obtain the following expression for the odds:
\begin{eqnarray}
\odds&=&\prod_{i=1}^{N_{\mathrm{SN}}}\left\{(1-\epsilon) + 
\epsilon\,\frac{1}{R\tau_i}\,\sum_{j=1}^{N_i}
\frac{\exp\left[(\yij\cdot\exi-1)/\sij^2\right]}
{\frac{1}{2}\sij^2\left(1-e^{-2/\sij^2}\right)}\right\}\nonumber\\
&\equiv&\prod_{i=1}^{N_{\mathrm{SN}}}\odds_i.
\label{odds2}
\end{eqnarray}

Some of the properties of this expression for $\odds$ are worth pointing
out: 

The term proportional to the average temporal exposure $\epsilon$
contains  a sum over candidate GRB counterparts to the SN.  Each term
in the sum consists of an exponential term that can penalize a
candidate GRB counterpart for excessive angular distance from the
position of the SN, and a denominator that can reward a candidate
counterpart for having a small error circle.  Thus, a GRB with a small
error circle that is not far from the position of the SN can produce a
large term in the sum.  A GRB whose error circle is very far from the
SN will produce an inconspicuous term in the sum, as will a GRB with a
very large error circle, irrespective of its position.

The term proportional to $\epsilon$ is also inversely proportional to
$R\tau_i\equiv\bar{N}_i$, the expected number of GRBs observed by BATSE
during the interval $\tau_i$.  This term prevents the expression for
$\odds_i$ from becoming large as a consequence of a large $\bar{N}_i$
resulting in one or more GRBs coinciding with the SN position purely by
chance.  In fact, we see that each term in the sum is inversely
proportional to $\bar{N}_i\sij^2/2$, a quantity that estimates the
number of bursts whose error circles bracket the SN by chance.

Finally, there the term $(1-\epsilon)$.  This term is an ``escape
hatch'', allowing for the possibility that all of the candidate GRB
counterparts are terrible fits because the actual counterpart was
missed due to incomplete temporal exposure.  This term has an
interesting consequence:  while a well-correlated individual SN-GRB
pair may produce a large $\odds_i$, the lack of such a pair cannot in
general produce a tiny $\odds_i$ if $\epsilon$ is not close to unity.
Thus, no \emph{individual} SN can rule out the hypothesized
association.  However, $H_1$ may still be convincingly be ruled out if
we have a collection of many SNe, the great majority of which have no
plausible GRB counterpart, since in that case we will have
$\odds\approx(1-\epsilon)^{N_{\mathrm{sn}}}$, which can be small.

The interpretation of the numerical value of $\odds$ is
straightforward: if $\odds\gg 1$ then the evidence favors $H_1$.  If
$\odds\ll 1$, the evidence favors $H_2$.  If $\odds\sim 1$, then the
evidence is insufficient to make a decisive case either way.

Note that $\odds\rightarrow 1$ as $\epsilon\rightarrow 0$.  In other
words, in the limit of no GRB observations at all, the evidence becomes
insufficient to budge the odds from the assigned prior value
$P(H_1|I)/P(H_2|I)=1$.  A similarly plausible limiting behavior of the
odds is $\lim_{\tau_i\rightarrow\infty}\odds_i\approx 1$, which may be
derived by considering the expected number of GRBs whose error circles
bracket the SN by chance in the long run.  Thus, in the limit of a
total lack of knowledge about the epoch of the SN explosion,  the
proliferation of candidate GRB counterparts introduces noise that
swamps our ability to distinguish between the two hypotheses.

It is also worth pointing out that this expression for the odds bears
some resemblance to the odds favoring the association of GRBs with host
galaxies derived by Band \& Hartmann (1998).  That work also compared
two ``simple'' hypotheses --- either GRBs have (intensity-redshift
correlated) host galaxies, or they don't.  Their expression for the
odds (equation [5] of Band \& Hartmann 1998) bears a structural
resemblance to our equation (\ref{odds2}), including a sum over
possible counterpart galaxies and a term accounting for the possibility
that the host galaxy was not observed because its luminosity was below
the detection threshold.  The main difference is that in their study,
GRBs play the role that SNe play in ours, with their GRB error circles
replacing our uncertainty in the time of the SN explosion, and with the
fraction of galaxies above the detection threshold replacing the BATSE
exposure.

There are two useful generalizations of this method:  we can add
Interplanetary Network (IPN) annuli to the data when they are
available, and we can consider a more general model in which not all
SNe produce detectable GRBs.

\subsection{IPN Annuli}

The inclusion of IPN annuli in the data is straightforward.  The IPN
catalog (\cite{hurley98}) gives the orientation of the line joining two
burst-detecting spacecraft, the angle $\theta$ between this direction
and the direction to the burst (which is the angular radius of the IPN
annulus), and a 3-$\sigma$ error in this angle, which we denote by
$\alpha$.  Thus $\theta$ is the angular radius and $\alpha$ is the
angular width of the IPN annulus.   Analogously, we define $\theta_{i}$
as the angle between the line joining two burst-detecting spacecraft
and the $i$th SN, and $\theta_{ij}$ as the angle between the line
joining two burst-detecting spacecraft and the $j$th burst possibly
associated with this SN.  We assume that under $H_1$, $\cos\theta_{ij}$
has a Gaussian distribution with mean $\cos\theta_i$ and error
$\sigma_{ij}=\sin\theta_{ij}\,\alpha_{ij}/3$ (with $\alpha_{ij}$ in
radians).  We also assume that under $H_2$, $\cos\theta_{ij}$ is
distributed uniformly in the range $[-1,1]$.  When an IPN annulus is
available, we replace the BATSE position by the annulus in the data
set. It is not difficult to show that the odds for an individual SN
then become
\begin{equation}
\odds_i=(1-\epsilon) + 
\epsilon\,\frac{1}{R\tau_i}\,\sum_{j=1}^{N_i} L_{ij},
\label{odds3}
\end{equation}
where
\begin{equation}
{\renewcommand{\arraystretch}{2.0}
L_{ij} = \left\{ 
\begin{array}{c@{\quad;\quad}l}
\frac{\displaystyle\exp\left[
-(\cos\theta_{ij}-\cos\theta_i)^2/2\sigma_{ij}^2
\right]}{\displaystyle\sqrt{2\pi}\sigma_{ij}/2}
&
\mbox{IPN annulus available,}
\\
\frac{\displaystyle\exp\left[(\yij\cdot\exi-1)/\sij^2\right]}
{\displaystyle\frac{1}{2}\sij^2\left(1-e^{-2/\sij^2}\right)}
&
\mbox{Otherwise}
\end{array}\right.
\label{li}
}
\end{equation}
The overall odds are still given by the product of the individual odds.

\subsection{More Complicated Model}

We may generalize $H_1$ to a model $H_1^\prime$ in which only a
fraction $f$ of SNe produce observable GRBs.  This may be due to
beaming, or for  some other reason.  The necessary modification of the
above formulas is straightforward.  We assume a uniform prior density
for $f$ in the range $[0,1]$.  The odds favoring model $H_1^\prime$
over model $H_2$ are denoted by $\odds^\prime$, and given by the
expression
\begin{equation}
\odds^\prime=\int_0^1df\,\odds(f),
\label{oprime}
\end{equation}
where
\begin{equation}
\odds(f)\equiv\prod_{i=1}^{N_{\mathrm{SN}}}
\left\{
(1-\epsilon f) + 
\epsilon f\,\frac{1}{R\tau_i}\,\sum_{j=1}^{N_i} L_{ij}
\right\}.
\label{of}
\end{equation}
In other words, $\odds(f)$ is constructed by setting the effective
probability of observing a GRB associated with a SN to $\epsilon f$,
rather than to $\epsilon$.

The quantity $\odds^\prime$ can help us decide whether  hypothesis
$H_1^\prime$ or $H_2$ is favored by the evidence.  If we were to find
that $H_1^\prime$ is strongly favored, we could then attempt to
estimate likely values for $f$.  For this purpose, we may use the
posterior probability density for $f$, given by
\begin{equation}
P(f|H_1^\prime,I)=\odds(f)/\odds^\prime
\label{pdf}
\end{equation}
We may construct a point estimate for $f$ by locating the maximum of
$P(f|H_1^\prime,I)$, and we may obtain interval estimates for $f$ by
finding intervals that contain a prescribed amount of probability ---
68\%, say --- as calculated by integrating $P(f|H_1^\prime,I)$.

The quantity $\odds^\prime$ is subject to an ambiguity:  it is
dependent upon our choice of prior probability density for $f$.  If
instead of a uniform prior density for $f$ in the range $[0,1]$ we had
chosen, for example, a uniform prior density in the range $[0,f_{\rm
SN}^0]$
(with $0<f_{\rm SN}^0<1$), then the expression for $\odds^\prime$ given in
Equation (\ref{oprime}) would be increased by a factor of ${f_{\rm SN}^0}^{-1}$.
Thus, a model that predicts small values of $f$ might find the
comparison with data less damaging than a model that is agnostic about
the value of $f$.

However, with $P(f|H_1^\prime,I)$ in hand, we may, if we wish, take a
different approach to the assessment of the plausibility of
$H_1^\prime$.  Instead of calculating the odds, we may calculate a
$3-\sigma$ upper bound for $f$.  The dependence of this upper bound on
the number of SNe in the sample may be calculated approximately
as follows:

Assuming $f$ is in fact small, so that not many coincidences are
observed, then the dependence of $P(f|H_1^\prime,I)$ on $f$ is seen
from Equation (\ref{of}) to be approximately
\begin{equation}
P(f|H_1^\prime,I)\sim (1-\epsilon f)^{N_{\mathrm{SN}}}.
\label{pfdep}
\end{equation}
After normalizing this expression, we may integrate it to produce the
cumulative probability:
\begin{eqnarray}
Q(f_{\rm SN}^0)&=&\int_0^{f_{\rm SN}^0} df\,P(f|H_1^\prime,I)\nonumber\\
&\approx& 
\frac{1-(1-\epsilon f_{\rm SN}^0)^{N_{\mathrm{SN}}+1}}
     {1-(1-\epsilon)^{N_{\mathrm{SN}}+1}}.
\label{qf}
\end{eqnarray}

\begin{figure}
\label{fqplot}
\plotone{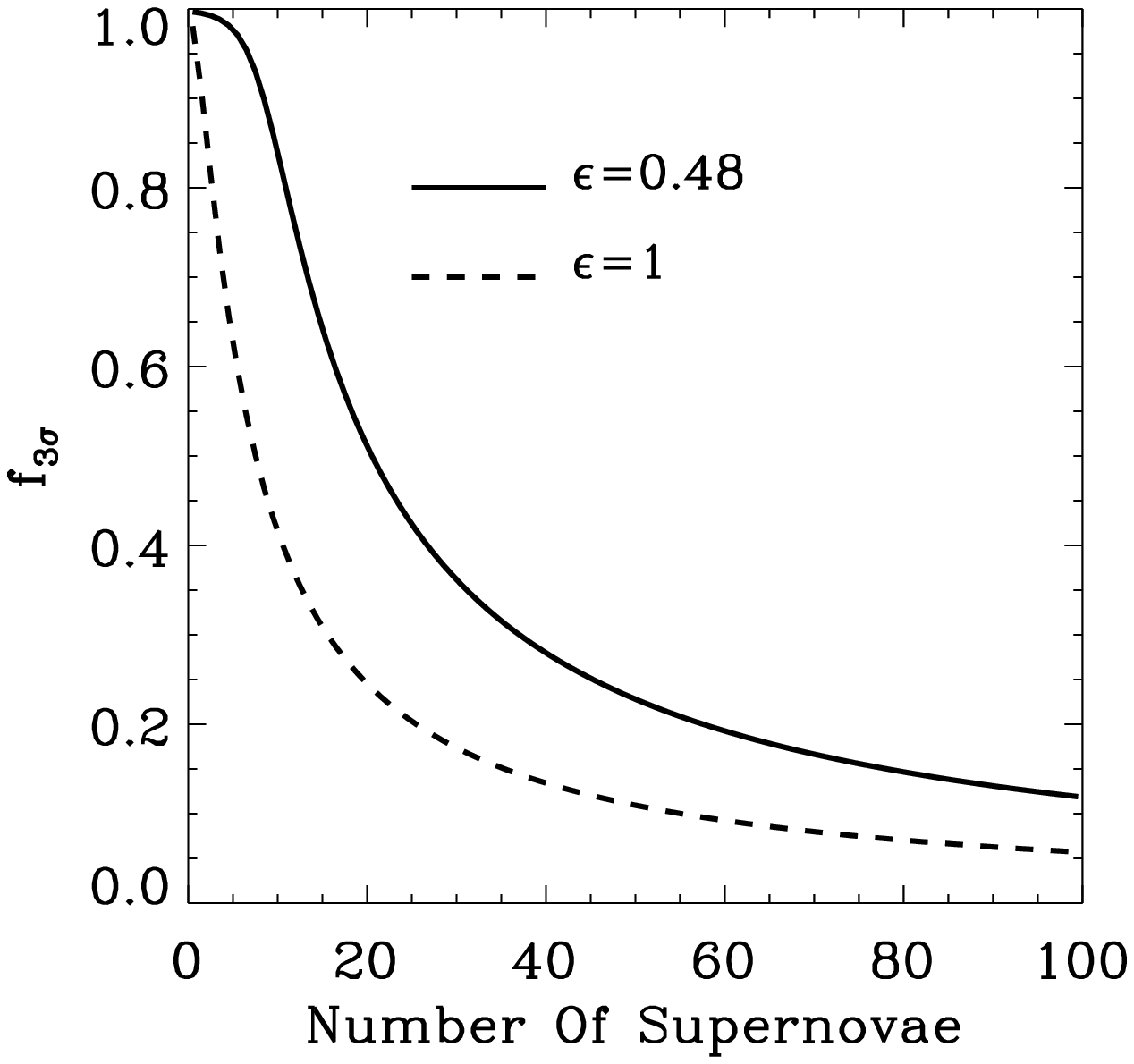}
\caption{
Dependence of the $3\sigma$ upper limit for $f$ on the number of
observed supernovae, assuming no excess of GRB-SN associations is
observed above what is expected by chance.  The upper curve was
calculated using the average BATSE exposure $\epsilon=0.48$
(\cite{hakkila98}), while the lower curve was calculated assuming
complete exposure.
}
\end{figure}

The quantity $Q(f_{\rm SN}^0)$ is the significance level of our upper
limit, say 99.73\%.  We may solve Equation (\ref{qf}) for $f_{\rm
SN}^0$, obtaining
\begin{equation}
f_{\rm SN}^0=\frac{1}{\epsilon}
\left\{1-\left[1-Q\times\left(1-(1-\epsilon)^{N_{\mathrm{sn}}+1}
         \right)\right]^{1/N_{\mathrm{SN}}+1}\right\}.
\label{fq}
\end{equation}
The dependence of $f_{\rm SN}^0$ on $N_{\mathrm{SN}}$ is plotted in
Figure \ref{fqplot}.  It is evident from the figure that even assuming
maximum exposure, the 3-$\sigma$ upper limit on $f$ can only be
expected to decrease very slowly with $N_{\mathrm{SN}}$.  It is
straightforward to show that for large $N_{\mathrm{SN}}$, the behavior
of $f_{\rm SN}^0$ is $f_{\rm SN}^0 \approx
-\ln(1-Q)/\epsilon(N_{\mathrm{SN}}+1)$.  Given the form of this
dependence on $N_{\mathrm{SN}}$, and given the relatively low rate
($\sim 10$ yr$^{-1}$) at which Type Ib-Ic SNe are currently being
discovered, it does not seem likely that observational evidence can
constrain $f$ in a significant manner anytime soon.

\section{Results}

We now apply the above methodology to the question of whether or not
the odds favor the hypothesis that a particular class of Type I SNe
produce GRBs over the hypothesis that they do not.  We first discuss
the samples of GRBs and Type I SNe that we use to address this
question.

\subsection{GRB and SN Samples}

The sample of GRBs that we use in our analysis consists of the BATSE 4B
catalog (\cite{meegan98},
http://www.batse.msfc.nasa.gov/data/grb/4bcatalog/), and BATSE bursts
that occurred subsequent to the 4B catalog but before 1 May 1998
(http://www.batse.msfc.nasa.gov/data/grb/catalog/).  The BATSE 4B
catalog consists of 1637 bursts, while the online archive contains an
additional 497 bursts through 1 May 1998.  We also use the Ulysses
supplement to the BATSE 4B catalog, which contains 219 BATSE bursts for
which 3rd IPN annuli have been determined (\cite{hurley98}).  Hurley
(private communication, 1998) has kindly made available at our request
3rd IPN annuli for an additional 9 BATSE bursts that occurred
subsequent to the period of the BATSE 4B catalog but before 1 May 1998.

We have compiled three Type I SNe samples.  The first is a sample of 37
Type Ia SNe (see Table 1) at low redshift ($z < 0.1$).  The data for
most of these events were kindly provided to us by the CfA SN Search
Team (Riess 1998, private communication).  The second is a sample of
46 moderate redshift ($0.1 < z < 0.830$) Type Ia SNe (see Table 2). 
The Supernova Cosmology Project (SCP) kindly supplied the data for
nearly all of these events (Perlmutter 1998, private communication). 
The third sample consists of 20 Type Ib, Ib/c, and Ic SNe (see Table
3).  We have compiled the data for these last events from information
available in the IAU Circulars and in various SNe catalogs (see the
footnotes to Table 3).  The procedure we use to estimate the range of
possible explosion dates $\Delta T$ for each SN event depends on the
type of SN and on the information available.  

For the low-redshift Type Ia SNe, estimate the explosion date using the
formula:   
\begin{equation} 
T = T_{\rm max} - 18.8\mbox{d} \times (1+z), \end{equation} 
where $T_{\rm max}$ is the estimated or observed date of maximum light
and $z$ is the redshift of the SN.  The uncertainty in the date of
maximum light is taken to be $\pm$1d for the CfA data supplied by
Riess.  In the sixteen cases where the date of maximum light has been
interpreted from spectra of the SN, we assign a greater uncertainty to
the date of maximum light.  When the language associated with the
spectral dating describes the observation as having been made near
maximum light, we adopt $\pm$6d for the uncertainty in the date of
maximum light.  We expand the uncertainty to $\pm$10d when the date of 
maximum light is estimated to be more than ten days prior to the date
on which the spectrum was taken or when the language of the Circular
suggests additional uncertainties.  We reject events for which our
evaluation of the uncertainty in the date of maximum light exceeds
$\pm$10d.   The total uncertainty in the explosion date that we assign
is the linear sum of the uncertainty in the date of maximum light and
an additional $\pm$2d for the uncertainty in the rise time predicted by
Type Ia explosion models.  The range of possible explosion dates
$\Delta T$ is given by adding and subtracting the total uncertainty
to/from the estimated explosion date $T$.

For the moderate $z$ Type Ia sample, the data provided by Perlmutter
include the SCP's best estimate of the explosion date, which was
computed using the formula:
\begin{equation}
T = T_{\rm max} - 18.8\mbox{d} \times (1+z) \times s,
\end{equation}
where $T_{\rm max}$ and $z$ are the same as before and $s$ is stretch 
factor, determined from the rate of decline of the Type Ia light curve 
and applied to the rising light curve (\cite{Perlmutter1}).  The 
uncertainty assigned by the SCP to the explosion date is $\pm$2.5d,
which we have rounded up to $\pm$3d for simplicity (see Table 2).  

Four additional SNe events are included in this sample.  The data for
these are taken from the Circulars and the estimated explosion dates
are calculated in the same way as were those for the low-$z$, Type Ia
sample.

For the sample of Type Ib, Ib/c, and Ic SNe, when it was possible to
estimate the date of maximum light, we calculated the estimated
explosion date using the formula:
\begin{equation}
T = T_{\rm max} - T_{\rm rise} \times (1+z),
\end{equation}
where $T_{\rm max}$ and $z$ are the same as before, and $T_{\rm rise}$
is taken to be 15d for Type Ib, 13d for Type Ib/c, and 12d for Type Ic
events.  The uncertainty in the date of maximum light and the range of
possible explosion dates $\Delta T$ are found by the same procedure as
for the low-$z$, Type Ia sample. 

When this method yields a range of possible explosion dates that
extends beyond the discovery date of the SN event, we take the end of
the range to be the discovery date.  Similarly, we limit the beginning
of the range of possible explosion dates when it extends to a date
earlier than the latest date on which an image was taken that does not
show the SN event.  We make the conservative assumption the the explosion
occurred no earlier than two days prior to the image date.  This
two-day allowance provides for the possibility that the brightness of
the SN may have been less than the limiting magnitude of the
observation on the date the image was taken (see Table 3).

There are three SNe in Table 3 (SN 1997C, 1998T, and 1998cc) for which
reliable information about the date of maximum light was unavailable. 
In these three cases, we were able to use other information to estimate
the range of possible explosion dates.  

For SN 1997C, we take the earliest possible explosion date to be two
days prior to the date of an image that shows no evidence of the SN.  
As before, the two-day allowance provides for the possibility that the
brightness of the SN may have been less than the limiting magnitude of
the observation on the date the image was taken.  There is also a
spectrum of this SN which shows it to be a Type Ic event 21 to 29 days
past maximum light (\cite{Li1}).  As a conservative estimate, we use
that date as the latest possible date of maximum light, and subtract 14
days ($T_{\rm rise}$ = 12d for a Type Ic SN plus 2d for the uncertainty in
the rise time predicted by explosion models), to find the latest
possible explosion date.

SN 1998T increased in brightness between two successive photometric
observations (\cite{Li2}), while a spectrum taken of SN 1998cc showed
features indicating that the SN had not yet reached maximum light 
(\cite{Jha2}).  Again, we are conservative and take the date of these 
observations to be the earliest possible date of maximum light. 
Consequently, the earliest possible explosion date for each SN is 17
days prior to the pre-maximum observation ($T_{\rm rise}$ = 15d for a
Type Ib SN plus 2d for the uncertainty in the rise time predicted by
explosion models). The latest possible explosion date for these events
is taken to be the discovery date (see Table 3).

Our sample of twenty Type Ib, Ib/c and Ic SNe includes five of the six 
SNe considered by \markcite{wang98}Wang \& Wheeler (1998); the sixth
(SN 1992ad) is a Type II SN (\cite{mcnaught92,filippenko92}) that was
mis-classified as a Type Ic SN by \markcite{wang98}Wang \& Wheeler.  In
two of the remaining five cases, the range of possible explosion dates
$\Delta T$ that we derive agrees closely with theirs; in the other
three cases they do not.  In the case of SN 1996N, the beginning of the
range of possible explosion dates that we adopt is similar to that of
\markcite{wang98}Wang \& Wheeler (1998), but the end of the range is
two weeks later.  In the case of SN 1997ei, the range of possible
explosion dates that we adopt is of the same duration as that of
\markcite{wang98}Wang \& Wheeler (1998), but shifted later by one
month.  The differences between our and \markcite{wang98}Wang \&
Wheeler's ranges of possible explosion dates for SN 1996N and SN 1997ei
do not affect our results.  However, in the third case (SN
1997X), our range of possible explosion dates begins 14$^{\rm d}$ later
than theirs because it is limited by the existence of an image that
shows no evidence of the SN (\cite{Nakano1}); as a result, the GRB that 
they associate with SN 1997X is excluded by the revised range of
possible explosion dates, and this makes a modest difference in our
results (see below).  

Finally, \markcite{wang98}Wang \& Wheeler (1998) list SN 1997ef in
their Table 1, but do not classify it and therefore do not
include it in their analysis.  We are able to classify it as a Type Ic
SN (\cite{Iwamoto2,Garnavich1}) and we include it in our analysis (see
Tables 3 and 4); the range of possible explosion dates we derive for
this SN is shifted later by about one month relative to that given by
\markcite{wang98}Wang \& Wheeler (1998).

\subsection{Type Ia Supernovae}

We first apply our methodology to the 83 events in our sample of Type
Ia SNe.  Since it is not expected that Type Ia SNe can produce GRBs,
these events constitute a ``control'' sample.  The results we find for
this sample illustrate the power of the methodology.  We find
overwhelming odds ($10^{21}:1$) against the hypothesis that all Type Ia
SNe produce observable GRBs (see Table 5).  Dividing our sample of Type
Ia SNe into two subsamples, a low-$z$ ($z \le 0.1$) subsample and a
moderate-$z$ ($z > 0.1$) subsample, we find overwhelming odds against
the hypotheses either that all low-$z$ Type Ia SNe or all moderate-$z$
Type Ia SNe produce observable GRBs (again, see Table 5).

We also find large odds ($34:1$) against the hypothesis that
some fraction $f_{\rm SN}$ of Type Ia SNe produce observable GRBs
(see Table 5).  Again dividing our sample of Type Ia SNe into two
subsamples, a low-$z$ ($z \le 0.1$) subsample and a moderate-$z$ ($z >
0.1$) subsample, we find moderate odds against the hypotheses  that
some fraction $f_{\rm SN}$ of either low-$z$ Type Ia SNe or
moderate-$z$ Type Ia SNe produce observable GRBs (again, see Table 5).

These results are not unexpected, given that an association between
Type Ia SNe is deemed unlikely on theoretical grounds and no
observational evidence has been reported linking the two.  Thus, the
Type Ia SNe constitute a control sample, validating the methodology we
have developed, and illustrating that, even with a BATSE mean temporal
exposure efficiency of $\epsilon = 0.48$, a SNe sample of moderate size
is sufficient to provide a severe test of the hypothesis that all Type
Ia SNe produce observable GRBs, and a strong test of the hypothesis 
that a fraction $f_{\rm SN}$ do.

\subsection{Type Ib, Ib/c and Ic Supernovae}

Applying our methodology to our sample of 20 Type Ib-Ic SNe, we find
very strong odds ($6000:1$) against the hypothesis that all Type Ib-Ic
SNe produce observable GRBs (see Table 5).  We find modest odds ($6:1$)
against the hypothesis that some fraction $f_{\rm SN}$ of Type Ia SNe
produce observable GRBs (see Table 5).  If we nevertheless assume that
this hypothesis is correct, we find that the fraction $f_{\rm SN}$ of
Type Ib, Ib/c and Ic SNe that produce observable GRBs must be less than
0.17, 0.42, and 0.70 with 68\%, 95\%, and 99.6\% probability,
respectively.  These limits are relatively weak because of the modest
size (20 events) of our sample of Type Ib-Ic SNe.

In order to verify that our results are insensitive to the range of
possible explosion dates  $\Delta T$ that we have derived, we repeated
our analysis of the  sample of Type Ib, Ib/c and Ic SNe with $\Delta T$
increased by $\pm$ 1, 2, 3, 5, and 10 days, except when the beginning
of the range is limited by  an image that does not show the SN or the
end of the range is limited by the SN discovery date.  The resulting
odds vary little (see Figure 1), demonstrating the robustness of our
conclusions.

\section{Discussion}

\subsection{Implications of Our Results}

We have applied a methodology based Bayesian inference to a sample of
83 Type Ia SNe.  We find overwhelming odds against the hypothesis that
all Type Ia SNe produce observable gamma-ray bursts, irrespective of
whether the SNe are at low- or high-redshift, and large odds against
the hypothesis that a fraction of Type Ia SNe produce observable
GRBs.  

Applying this methodology to a sample of 20 Type Ib-Ic SNe, we find
very large odds against the hypothesis that all Type Ib-Ic SNe produce
observable GRBs, and modest odds against the hypothesis that some
fraction $f_{\rm SN}$ of Type Ia SNe produce observable GRBs (see Table
5).  If we nevertheless assume that this hypothesis is correct, we find
that the fraction $f_{\rm SN}$ of Type Ib, Ib/c and Ic SNe that produce
observable GRBs is less than 0.70 with 99.7\% probability.  This limit
is relatively weak because of the modest size (20 events) of our sample
of Type Ib-Ic SNe.

Type Ib, Ib/c and Ic SNe are now being found at a rate of about eight a
year, so that the size of the sample of known Type Ib-Ic SNe should
double within three years and triple within about five years.  One
might hope that future analyses, using the statistical methodology that
we have presented here, could either show that the association between
Type Ib-Ic SNe and GRBs is rare, or confirm the proposed association. 
Unfortunately, equation (18) and Figure 2 show that achieving the
former will be difficult: the limit on the fraction $f_{\rm SN}$ of
Type Ib-Ic SNe that produce observable GRBs scales like $N_{\rm
SN}^{-1}$ for large $N_{\rm SN}$, and therefore tripling the size of
the sample of known Type Ib-Ic SNe without observing an additional
possible SN -- GRB association would only reduce the 99.7\% probability
upper limit on $f_{\rm SN}$ to 0.24.

An alternative approach is to use the upper limit on the fraction of
BATSE bursts that can have come from a homogeneous, isotropic
distribution to place an upper limit on the fraction of nearby Type
Ib-Ic SNe that can have produced observable GRBs.  We infer from
\markcite{smith93}Smith \& Lamb (1993) that no more than roughly 20\%
of the BATSE burst can have come from such a population, corresponding
to $\approx 400$ bursts during the $\approx 7$ years covered by our
study.

GRB 980425 had a 1024 msec peak flux $F_{\rm peak}^{1024} = 0.9$
photons cm$^{-2}$ s$^{-1}$, which is less than a factor of four above
the BATSE threshold.  Consequently, assuming that the association
between SN 1998bw and GRB 980425 is real and that the GRBs produced by
Type Ib-Ic SNe are standard candles, BATSE only detects Type Ib-Ic SNe
that lie at redshifts smaller than $z \approx 2 z_{\rm SN 1998bw} =
0.016$.  This is a sampling distance of $\sim 48 h^{-1}$~Mpc.

Given a supernova rate of $\sim2$ per $L_\star$ galaxy per century
(\cite{strom95}), a density of $L_\star$ galaxies of $0.01
h^3$~Mpc$^{-3}$, and that approximately 2/7 of these are SNe of Types
Ib, Ib/c and Ic (\cite{woosley86}, \cite{strom95}), we find that during
the $\approx 7$ years covered by our study, the number of SNe that
occurred and could have produced GRB detectable by BATSE is 185.  Given
the BATSE average temporal exposure $\epsilon = 0.48$, this implies
that at most $\sim 90$ such SNe could have been detected by BATSE. 
Comparing this number with the number $\approx 400$ of GRBs that can
have come  from such a population, we derive an upper limit on the
fraction of Type Ib-Ic SNe that can have produced an observable GRB of
$f_{\rm SN}^{F_{\rm peak}} \approx 400/90 > 1$. This means that no
constraint may be placed on $f_{\rm SN}^{F_{\rm peak}}$ by this method,
and shows that the method used in the present study, as opposed to
modeling of the BATSE angular and brightness distributions, provides
the more stringent constraint. 

One can also approach the proposed association between SNe and GRBs
from the opposite direction.   The interesting question, from this
point of view, is what fraction $f_{\rm GRB}$ of the GRBs detected by
BATSE could have been produced by Type Ib-Ic SNe?  The above discussion
indicates that  $f_{\rm GRB}$ can be no more than $f_{\rm GRB}^0
\lesssim (90/2000) f_{\rm SN}^0 = 0.045 f_{\rm SN}^0 \lesssim 0.03$,
where in the last step we have used the 99.7\% probability
upper limit derived from our Bayesian analysis.  

This question can also be addressed by examining relatively accurate
($\leq 1'$) GRB positional error circles for the presence of Type
Ib-Ic SNe.  BeppoSAX observations have already placed a weak limit on
this fraction:  Setting aside GRB 980425, none of the remaining 14
BeppoSAX WFC GRB error circles has been found to contain a Type Ib-Ic
SN.  The HETE-II mission is expected to place somewhat stronger limits
on this fraction (or possibly confirm the proposed association between
GRBs and Type Ib-Ic SNe), since it is expected to provide a larger
number of relatively accurate positions for GRBs.

The limit $f^0_{\rm GRB}$ that can be placed on the fraction $f_{\rm
GRB}$ of GRBs that can have been produced by SNe is given by equation
(18), with the number of SNe, $N_{\rm SN}$, replaced by the number of
GRBs, $N_{\rm GRB}$, and $\epsilon$ set equal to one (we assume that
the efficiency of detecting a Type Ib-Ic SN in a relatively accurate
GRB positional error circle is 100\%).  Thus the dashed curve in Figure
2 shows this limit as a function of $N_{\rm GRB}$.  Equation (18) and
Figure 2 show that this limit scales as $N_{\rm GRB}^{-1}$ for large
$N_{\rm GRB}$, as one intuitively expects.  Consequently, placing a
99.7\% upper limit on $f_{\rm GRB}$ that is more stringent than the
rough limit derived above, or confirming the proposed association
between GRBs and Type Ib-Ic SNe if such associations are rare, will
require a mission that produces relatively accurate ($\leq 1'$)
positions for a very large number ($\geq 1000$) of GRBs.

\subsection{Comparison with Other Work}

\markcite{wang98}Wang \& Wheeler (1998) have reported that analysis
of a sample of six Type Ib-Ic SNe using an {\it a posteriori}
frequentist  statistic rules out at the $10^{-5}$ significance level
the null hypothesis that the positions on the sky of Type Ib, Ib/c, and
Ic SNe and GRBs are uncorrelated.  The statistic they used is the
probability that at least one of the BATSE positional error circles for
the GRBs occurring within the range of possible explosion dates $\Delta
T$ of a Type Ib-Ic SN includes the position of the SN.

In order to understand the apparent discrepancy between our results and
those of Wang \& Wheeler (1998), we have re-analyzed their sample of
Type Ib-Ic SNe, except for the elimination of SN1992ad (which is a Type
II SN, not a Type Ic as Wang \& Wheeler assumed), and corrections to
their ranges of possible explosion dates.  We use a generalization of
their methodology that is applicable to a sample in which not all of
the SNe are bracketed by a GRB error circle.  We have also applied
this  generalization of their methodology to our larger sample of 20
Type Ib-Ic SNe.

Following Wang \& Wheeler (1998), we ascribe to the $i$th SN a
probability $f_i$ that its position should be ``bracketed'' by chance
by at least one of the positional error circles of the GRBs that
occurred during  its range of possible explosion dates $\Delta T$.  We
assume the power-law model of BATSE systematic errors
(\cite{graziani96}) and combine these systematic errors with the BATSE
statistical errors to produce a total 1-$\sigma$ error circle radius
for each burst.  We then multiply that error circle radius by 3 to
produce the bracketing circle radius $\mu_i$.  Note that $\mu$ is
\emph{larger} than the ``3-$\sigma$'' (99.7\% probability) error circle
radius, which would be obtained by multiplying the 68.3\% radius by
2.27 - as may be inferred from the $\chi^2$ distribution with two
degrees of freedom.  We multiply by 3 in order that our procedure
agree as closely as possible with that of Wang \& Wheeler (1998).  

Following Wang \& Wheeler (1998), the $f_i$ are given by
\begin{equation}
f_i=1-\prod_{j=1}^{N_i}\left[1-\frac{1}{2}(1-\cos\mu_i)\right].
\label{fi}
\end{equation}

Wang \& Wheeler (1998) asserted that the positions of all six of the
SNe in their sample were bracketed by the positional error circles of
at least one GRB, and employed a probability that was simply the
product of all their $f_i$.  In the more general case in which not all
the SN are bracketed, one cannot apply their procedure, or even merely
multiply by $f_i$ for each bracketed SN and by $1-f_i$ for each
non-bracketed SN.  The reason this latter procedure fails is that it is
guaranteed to produce a small number --- irrespective of whether or not
many SN were bracketed --- because the product of many numbers that lie
between zero and one can be small, even if most of them lie near one.  

The problem is that by calculating the probability that \emph{this} SN
should be bracketed and \emph{that} one should not be, we are inquiring
after a peculiar state of the data, rather than a generic one.  The
distinction is analogous to the distinction between microstates and
macrostates in statistical mechanics.  We should instead calculate the
probability of some generic feature of the observed data.  We choose
the number $N_b$ of bracketed SN as our generic property, and calculate
the probability that the observed number of bracketed SN should have
been $N_b^{(\mathrm{obs})}$ or higher.  Obviously, there are many
contributing configurations
$\mathcal{C}(N_b)\equiv\{q_i;i=1,\ldots,N_{\mathrm{sn}}\}$ in which
$N_b$ of the $q_i$ assume the values $f_i$, and the remainder assume
the value $1-f_i$.  We must sum over the contribution of all such
configurations.  Our significance is then given by
\begin{equation}
S=
\sum_{N_b=N_b^{(\mathrm{obs})}}^{N_{\mathrm{sn}}}\,
\sum_{\mathcal{C}(N_b)}\,
\prod_{i=1}^{N_{\mathrm{sn}}} q_i.
\label{signif}
\end{equation}

For the modified Wang \& Wheeler sample (SN1994I, SN1996N, SN1997X,
SN1997ei, and SN1998T), we find that all but SN1997X are bracketed by a
GRB in the explosion time window.  The resulting significance is
$S=2.6\%$.  By comparison, Wang \& Wheeler (1998) found $S=1.5\times
10^{-5}$.  For the full sample of Type Ib-Ic SNe, we find 9 bracketed
SNe out of 20, with a significance $S=35\%$. 

Thus, upon re-analyzing the corrected sample of Type Ib-Ic SNe studied
by  \markcite{wang98}Wang \& Wheeler (1998), we find no significant
evidence for an association between GRBs  and Type Ib, Ib/c and Ic
SNe.  Moreover, using a sample of Type Ib, Ib/c and Ic SNe that is four
times larger, we find that the significance becomes even weaker --- not
stronger, as would be expected if the association were real.  We
conclude that a frequentist analysis similar to that performed by
\markcite{wang98}Wang \& Wheeler (1998) shows no evidence for an
association between Type Ib-Ic SNe and GRBs, consistent with the very
strong evidence against such a correlation that we find from our
Bayesian analysis.

\markcite{kippen98}Kippen et al. (1998) have approached the proposed
association between SNe and GRBs from the opposite direction.  They
have asked the question, what fraction of BATSE bursts can have been
produced by {\it known} SNe?  They find no evidence of any correlation
between SNe and BATSE bursts, and derive a 3-$\sigma$ (99.7\%) limit on
any such fraction of 1.5\%, which corresponds to $\approx 18$ bursts. 
Unfortunately, this result is not very interesting because known SNe
comprise such a small fraction ($\lesssim 10^{-5}$) of the SNe that
occurred during the time interval they study.  This, together with our
earlier result that the fraction of BATSE bursts that could be produced
by SNe is $\lesssim$ 3\%, implies that a negligible fraction ($\sim
10^{-7}$) of BATSE bursts could be produced by {\it known} SNe. 
\markcite{kippen98}Kippen et al.'s result is further weakened by the
fact that their study does not distinguish among Type II, Type Ia, and
Type Ib-Ic SNe, whereas it is only the last type of SNe that are
thought possibly to produce GRBs.

\section{Conclusions}

We find very large odds against the hypothesis that all Type Ib-Ic SNe
produce observable GRBs, and moderate odds against the hypothesis that
a fraction of Type Ib-Ic supernovae produce observable GRBs.  We have
also re-analyzed a corrected version of the \markcite{wang98}Wang \&
Wheeler (1998) sample of Type Ib-Ic SNe, as well as our larger sample of
20 Type Ib-Ic SNe, using a generalization of their frequentist method. 
We find no significant evidence of a correlation between Type Ib-Ic SNe
and GRBs in either case, consistent with the very strong evidence
against such a correlation that we find from our Bayesian analysis.

While these statistical studies cannot address the question of whether
a particular GRB is produced by a particular Type Ib-Ic SNe (i.e., GRB
980425 by SN 1998bw), they show not only that there is no evidence of
an association between Type Ib-Ic SNe and GRBs, but that the odds
against the hypothesis that all Type Ib-Ic SNe produce observable GRBs
are very large, and the odds against the hypothesis that a fraction of
them do are moderate.  These results suggest that considerable caution
is warranted before accepting the association between GRB 980425 and SN
1998bw.  This is particularly the case, given that there exists in the
BeppoSAX WFC error circle a fading X-ray source whose behavior is
consistent with the power-law temporal decline observed for the X-ray
afterglows of other BeppoSAX bursts (\cite{pian98b}).  This fading
X-ray source might well be the counterpart to the GRB, rather than SN
198bw.

\acknowledgements 
We gratefully acknowledge the information on Type Ia SNe supplied to us
by Adam Riess and the CfA Supernova Group, and by Saul Perlmutter and
the Berkeley Supernova Group.  We also gratefully acknowledge the
information available to us via the Compton Gamma-Ray Observatory
Science Center and the information  provided by Kevin Hurley on the 3rd
IPN annuli  for GRBs that have occurred subsequent to the BATSE 4B
catalog.  We also thank Cole Miller and Jean Quashnock for helpful
conversations.

\newpage

\begin{figure}
\label{window}
\plotone{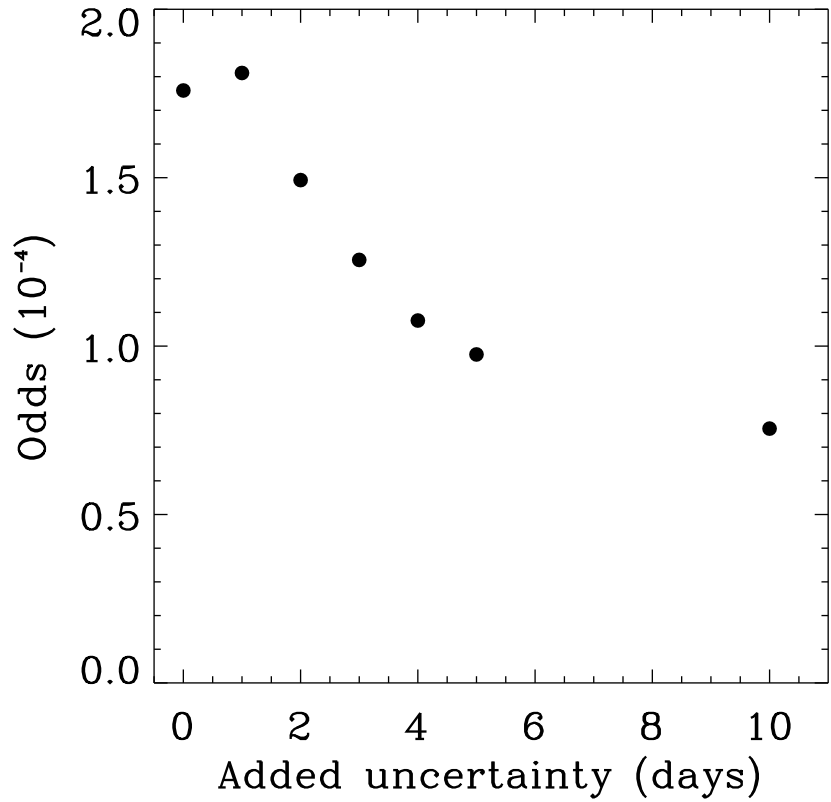}
\caption{
Variation of the odds favoring the association hypothesis $H_1$ with
added uncertainty in the time window of occurence of the supernova
explosion.  The added uncertainty is the number of days added (linearly)
to the half-width of the time window.
}
\end{figure}

\newpage
\begin{deluxetable}{lcccccc}
\tablewidth{500pt}

\tablecaption{\label{lowzIa}
 Estimated Explosion Dates and Positions for Low-z Type Ia Supernovae}
\tablehead{SN&z&Discovery&Max. Light&Range Exp. Dates&RA&DEC}

\startdata
1993B  & .0690 & 930117 & 930118 $\pm$6\tablenotemark{a} & 921221 $-$ 930106 & 10 34 51.38 & $-$34 26 30.0\nl
1993Q  & .0300 & 930528 & 930513 $\pm$10\tablenotemark{b}& 930412 $-$ 930506 & 20 35 46.94 & $-$42 47 33.4\nl
1993ac & .0493 & 931013 & 931010 $\pm$1\tablenotemark{c} & 930918 $-$ 930924 & 05 46 23.55 & +63 22 07.0\nl
1993ae & .0190 & 931107 & 931031 $\pm$1\tablenotemark{c} & 931009 $-$ 931015 & 01 29 48.92 & $-$01 58 37.2\nl
1994B  & .0899 & 940116 & 940123 $\pm$1\tablenotemark{d} & 931231 $-$ 940106 & 08 20 40.92 & +15 43 48.4\nl
1994C  & .0515 & 940305 & 940227 $\pm$1\tablenotemark{d} & 940205 $-$ 940211 & 07 56 40.27 & +44 52 19.3\nl
1994M  & .0230 & 940429 & 940430 $\pm$1\tablenotemark{c} & 940408 $-$ 940414 & 12 31 08.61 & +00 36 19.9\nl
1994Q  & .0290 & 940602 & 940526 $\pm$1\tablenotemark{c} & 940504 $-$ 940510 & 16 49 51.14 & +40 25 55.8\nl
1994S  & .0152 & 940604 & 940614 $\pm$1\tablenotemark{c} & 940523 $-$ 940529 & 12 31 21.86 & +29 08 04.2\nl
1994T  & .0347 & 940611 & 940607 $\pm$1\tablenotemark{c} & 940516 $-$ 940522 & 13 18 56.16 & $-$01 53 15.0\nl
1994U  & .0043 & 940627 & 940705 $\pm$1\tablenotemark{d} & 940614 $-$ 940620 & 13 04 56.13 & $-$07 56 51.5\nl
1994ae & .0043 & 941114 & 941128 $\pm$1\tablenotemark{c} & 941107 $-$ 941113 & 10 47 01.95 & +17 16 31.0\nl
1995D  & .0065 & 950210 & 950219 $\pm$1\tablenotemark{c} & 950129 $-$ 950204 & 09 40 54.75 & +05 08 26.2\nl
1995E  & .0116 & 950220 & 950224 $\pm$1\tablenotemark{c} & 950202 $-$ 950208 & 07 51 56.75 & +73 00 34.6\nl
1995M  & .0531 & 950422 & 950415 $\pm$1\tablenotemark{d} & 950324 $-$ 950330 & 09 38 41.78 & $-$12 20 07.9\nl
1995ac & .0500 & 950922 & 951002 $\pm$1\tablenotemark{c} & 950910 $-$ 950916 & 22 45 34.14 & $-$08 45 04.7\nl
1995ae & .0677 & 950922 & 950922 $\pm$1\tablenotemark{d} & 950850 $-$ 950922 & 23 16 55.65 & $-$02 04 36.4\nl
1995ak & .0230 & 951027 & 951030 $\pm$1\tablenotemark{c} & 951008 $-$ 951014 & 02 45 48.83 & +03 13 50.1\nl
1995al & .0051 & 951101 & 951107 $\pm$1\tablenotemark{c} & 951017 $-$ 951023 & 09 50 55.97 & +33 33 09.4\nl
1995bd & .0160 & 951219 & 960102 $\pm$1\tablenotemark{c} & 951211 $-$ 951217 & 04 45 21.24 & +11 04 02.5\nl
1996C  & .0296 & 960215 & 960212 $\pm$1\tablenotemark{c}& 960121 $-$ 960127 & 13 50 48.60 & +49 19 07.1\nl
1996X  & .0068 & 960412 & 960417 $\pm$1\tablenotemark{c}& 960327 $-$ 960402 & 13 18 01.13 & $-$26 50 45.3\nl
1996V  & .0247 & 960328 & 960330 $\pm$1\tablenotemark{d}& 960308 $-$ 960314 & 11 21 31.23 & +02 48 40.4\nl
1996Z  & .0076 & 960516 & 960513 $\pm$1\tablenotemark{c}& 960422 $-$ 960428 & 09 36 44.82 & $-$21 08 51.7\nl
1996ai & .0032 & 960616 & 960620 $\pm$1\tablenotemark{c}& 960530 $-$ 960605 & 13 10 58.13 & +37 03 35.4\nl
1996bk & .0068 & 961012 & 961009 $\pm$1\tablenotemark{c}& 960918 $-$ 960924 & 13 46 57.98 & +60 58 12.9\nl
1996bl & .0068 & 961011 & 961020 $\pm$1\tablenotemark{c}& 960929 $-$ 961005 & 00 36 17.97 & +11 23 40.5\nl
1996bo & .0173 & 961018 & 961030 $\pm$1\tablenotemark{c}& 961008 $-$ 961014 & 01 48 22.80 & +11 31 15.8\nl
1996bv & .0167 & 961103 & 961116 $\pm$1\tablenotemark{c}& 961025 $-$ 961031 & 06 16 13.00 & +57 03 08.9\nl
1996by & .0137 & 961214 & 961220 $\pm$6\tablenotemark{e}& 961125 $-$ 961207 & 05 58 24.96 & +68 27 12.1\nl
1996ca & .0167 & 961215 & 961220 $\pm$6\tablenotemark{e}& 961125 $-$ 961207 & 22 30 59.26 & $-$13 59 50.9\nl
1997Y  & .0162 & 970202 & 970209 $\pm$6\tablenotemark{e}& 970115 $-$ 970127 & 12 45 31.40 & +54 44 17.0\nl
1997bp & .0077 & 970406 & 970408 $\pm$1\tablenotemark{d}& 970318 $-$ 970324 & 12 46 53.75 & $-$11 38 33.2\nl
1997dg & .0340 & 970927 & 970930 $\pm$6\tablenotemark{e}& 970905 $-$ 970917 & 23 40 14.21 & +26 12 11.8\nl
1997dt & .0073 & 971122 & 971123 $\pm$6\tablenotemark{e}& 971030 $-$ 971111 & 23 00 02.93 & +15 58 50.9\nl
1998V  & .0176 & 980310 & 980304 $\pm$6\tablenotemark{e}& 980210 $-$ 980222 & 18 22 37.40 & +15 42 08.4\nl
1998bu & .0033 & 980509 & 980515 $\pm$6\tablenotemark{e}& 980421 $-$ 980503 & 10 46 46.03 & +11 50 07.1\nl
\enddata

\tablenotetext{a}{\cite{Phillips3}\ \
$^{b}$\cite{Della Valle1}\ \
$^{c}$Adam G. Riess 1998, private communication\ \ 
$^{d}$\cite{Riess1}\ \  
$^{e}$\cite{CfA1}}

\end{deluxetable}

\begin{deluxetable}{lcccccc}
\tablewidth{500pt}

\tablecaption{\label{hizIa}
 Estimated Explosion Dates and Positions for High-z Type Ia Supernovae}
\tablehead{SN&z&Discovery&Max. Light&Range Exp. Dates&RA(2000)&DEC(2000)}

\startdata
1992bi  & 0.458&   920421& na&   920313 $-$  920319\tablenotemark{a}  & 16 10 12.74&   +39 47  12.7\nl
1994F   & 0.354&   940109& na&   931226 $-$  940101\tablenotemark{a}  & 11 49 59.53&   +10 42  59.5\nl
1994G   & 0.425&   940213& na&   940124 $-$  940130\tablenotemark{a}  & 10 19 16.72&   +50 52  16.7\nl
1994H   & 0.374&   940108& na&   931225 $-$  931231\tablenotemark{a}  & 02 40 04.60& $-$01 34  04.6\nl
1994al  & 0.420&   940108& na&   931221 $-$  931227\tablenotemark{a}  & 03 06 22.41&   +17 18  22.4\nl
1994am  & 0.372&   951022& na&   940106 $-$  940112\tablenotemark{a}  & 02 40 02.06& $-$01 37  02.1\nl
1994an  & 0.378&   941031& na&   941005 $-$  941011\tablenotemark{a}  & 22 44 18.79&   +00 06  18.8\nl
1995K   & 0.478&   950330& 950401 $\pm1$\tablenotemark{b}&   950228 $-$  950308  & 10 50 47.00& $-$09 15  07.4\nl
1995ao  & 0.240&   951118& 951123 $\pm6$\tablenotemark{c}&   951023 $-$  951108  & 02 57 30.70& $-$01 41  19.8\nl
1995ap  & 0.300&   951118& 951123 $\pm6$\tablenotemark{c}&   951022 $-$  951107  & 03 12 28.13&   +00 41  43.4\nl
1995ay  & 0.480&   951120& na&   951030 $-$  951105\tablenotemark{a}  & 03 01 07.49&   +00 21  07.5\nl
1995aq  & 0.453&   951119& na&   951021 $-$  951027\tablenotemark{a}  & 00 29 04.22&   +07 51  04.2\nl
1995ar  & 0.497&   951119& na&   951028 $-$  951103\tablenotemark{a}  & 01 01 20.37&   +04 18  20.4\nl
1995as  & 0.498&   951119& na&   951019 $-$  951025\tablenotemark{a}  & 01 01 35.26&   +04 26  35.3\nl
1995at  & 0.655&   951120& na&   951024 $-$  951030\tablenotemark{a}  & 01 04 50.90&   +04 33  50.9\nl
1995aw  & 0.400&   951119& na&   951103 $-$  951109\tablenotemark{a}  & 02 24 55.50&   +00 53  55.5\nl
1995ax  & 0.615&   951119& na&   951020 $-$  951026\tablenotemark{a}  & 02 26 25.77&   +00 48  25.8\nl
1995az  & 0.450&   951120& na&   951107 $-$  951113\tablenotemark{a}  & 04 40 33.56& $-$05 30  33.6\nl
1995ba  & 0.388&   951120& na&   951017 $-$  951023\tablenotemark{a}  & 08 19 06.45&   +07 43  06.4\nl
1996aj  & 0.105&   960615& 960604 $\pm10$\tablenotemark{d}&   960502 $-$ 960526  & 13 29 06.82& $-$29 14  02.0
\tablebreak
1996cf  & 0.570&   960317& na&   960220 $-$  960226\tablenotemark{a}  & 10 48 50.96&   +00 03  51.0\nl
1996cg  & 0.460&   960317& na&   960215 $-$  960221\tablenotemark{a}  & 08 24 13.33&   +03 24  13.3\nl
1996ci  & 0.495&   960317& na&   960218 $-$  960224\tablenotemark{a}  & 13 45 56.16&   +02 26  56.2\nl
1996ck  & 0.656&   960317& na&   960216 $-$  960222\tablenotemark{a}  & 12 48 35.19&   +00 46  35.2\nl
1996cl  & 0.828&   960318& na&   960214 $-$  960220\tablenotemark{a}  & 10 56 59.13& $-$03 37  59.1\nl
1996cm  & 0.450&   960318& na&   960216 $-$  960222\tablenotemark{a}  & 15 30 11.25&   +05 55  11.3\nl
1996cn  & 0.430&   960318& na&   960227 $-$  960302\tablenotemark{a}  & 13 48 27.22&   +02 27  27.2\nl
1997F   & 0.580&   970105& na&   961221 $-$  961227\tablenotemark{a}  & 04 55 14.25& $-$05 51  14.2\nl
1997G   & 0.763&   970105& na&   961130 $-$  961206\tablenotemark{a}  & 04 58 30.21& $-$03 16  30.2\nl
1997H   & 0.526&   970105& na&   961205 $-$  961211\tablenotemark{a}  & 04 59 36.56& $-$03 09  36.6\nl
1997I   & 0.172&   970105& na&   961223 $-$  961229\tablenotemark{a}  & 04 59 37.30& $-$03 09  37.3\nl
1997J   & 0.619&   970105& na&   961205 $-$  961211\tablenotemark{a}  & 07 41 17.82&   +09 33  17.8\nl
1997K   & 0.592&   970106& na&   961205 $-$  961211\tablenotemark{a}  & 07 54 55.07&   +04 19  55.1\nl
1997L   & 0.550&   970105& na&   961219 $-$  961225\tablenotemark{a}  & 08 21 57.12&   +03 53  57.1\nl
1997N   & 0.180&   970105& na&   961130 $-$  961206\tablenotemark{a}  & 08 23 50.01&   +03 28  50.0\nl
1997O   & 0.374&   970106& na&   961223 $-$  961229\tablenotemark{a}  & 08 24 02.49&   +04 07  02.5\nl
1997P   & 0.472&   970106& na&   961206 $-$  961212\tablenotemark{a}  & 10 55 55.90& $-$03 56  55.9\nl
1997Q   & 0.430&   970106& na&   961207 $-$  961213\tablenotemark{a}  & 10 56 51.45& $-$03 58  51.4\nl
1997R   & 0.657&   970106& na&   961220 $-$  961226\tablenotemark{a}  & 10 57 19.20& $-$03 54  19.2\nl
1997S   & 0.612&   970106& na&   961202 $-$  961208\tablenotemark{a}  & 10 57 51.57& $-$03 45  51.6
\tablebreak
1997ac  & 0.320&   960317& na&   970131 $-$  970206\tablenotemark{a}  & 08 24 05.21&   +04 11  05.2\nl
1997af  & 0.579&   960317& na&   970221 $-$  970227\tablenotemark{a}  & 08 23 52.68&   +04 08  52.7\nl
1997ai  & 0.450&   970305& na&   970202 $-$  970208\tablenotemark{a}  & 10 48 57.62&   +00 31  57.6\nl
1997aj  & 0.581&   970305& na&   970218 $-$  970224\tablenotemark{a}  & 10 55 52.98& $-$03 59  53.0\nl
1997am  & 0.416&   970305& na&   970127 $-$  970202\tablenotemark{a}  & 10 57 31.52& $-$03 13  31.5\nl
1997ap  & 0.830&   970305& na&   970207 $-$  970213\tablenotemark{a}  & 13 47 09.90&   +02 23  09.9\nl

\enddata

\tablenotetext{a}{\cite{Perlmutter1}; and private communication\ \
$^{b}$\cite{Riess1}\ \
$^{c}$\cite{Kirshner1}\ \
$^{d}$\cite{Garnavich2}}

\end{deluxetable}

\begin{deluxetable}{llccllccc}
\newcommand{\ssz}{\footnotesize}

\tablewidth{500pt}
\tablecaption{\label{Ibc}
 Estimated Explosion Dates and Positions for SN Type Ib, Ic, \& Ib/c}
\tablehead{\ssz SN&\ssz Type&\ssz z$^\#$&\ssz Discovery&\ssz Max. Light&\ssz Range Exp. Dates
&\ssz W$^2$Range&\ssz RA(2000)&\ssz DEC(2000)}

\startdata
\ssz 1992ar&\ssz Ic&\ssz .1450&\ssz 920727&\ssz 920802$\pm$6 \tablenotemark{a}&\ssz 920712$-$920727*&\ssz na&\ssz 23 17 28.40&\ssz $-$44 38 53.8\nl

\ssz 1993P&\ssz Ic&\ssz .0480&\ssz 930518&\ssz 930521$\pm$6 \tablenotemark{b}&\ssz 930501$-$930517&\ssz na&\ssz 13 29 25.80&\ssz $-$30 24 47.4\nl

\ssz 1994I&\ssz Ic&\ssz .0015&\ssz 940402&\ssz 940411$\pm$2 \tablenotemark{c}&\ssz 940326$-$940402*&\ssz 940329$-$949402&\ssz 13 29 54.01&\ssz +47 11 31.7\nl

\ssz 1994ai&\ssz Ic&\ssz .0050&\ssz 941220&\ssz 941224$\pm$6 \tablenotemark{d}&\ssz 941204$-$941220&\ssz na&\ssz 02 23 06.17&\ssz $-$21 13 58.3\nl

\ssz 1995F&\ssz Ic&\ssz .0051&\ssz 950210&\ssz 950207$\pm$10 \tablenotemark{e}&\ssz 950114$-$950207&\ssz na&\ssz 09 04 57.40&\ssz +59 55 58.7\nl

\ssz 1996D&\ssz Ic&\ssz .0159&\ssz 960209&\ssz 960218$\pm$6 \tablenotemark{f}&\ssz 960129$-$960209*&\ssz na&\ssz 04 34 00.00&\ssz $-$08 35 00.0\nl

\ssz 1996N&\ssz Ib/c&\ssz .0047&\ssz 960312&\ssz 960309$\pm$10 \tablenotemark{g}&\ssz 960214$^\dag$$-$960308&\ssz 960210$-$960224&\ssz 03 38 55.31&\ssz $-$26 20 04.1\nl

\ssz 1996aq&\ssz Ic&\ssz .0055&\ssz 960817&\ssz 960820$\pm$6 \tablenotemark{h}&\ssz 960731$-$960816&\ssz na&\ssz 14 22 22.73&\ssz $-$00 23 24.3\nl

\ssz 1996cd&\ssz Ib/c&\ssz .0480&\ssz 961216&\ssz 961224$\pm$10 \tablenotemark{i \&\ssz  j}&\ssz 961129$-$961216*&\ssz na&\ssz 07 57 20.73&\ssz +11 12 23.7\nl

\ssz 1997B&\ssz Ic&\ssz .0104&\ssz 970114&\ssz 970104$\pm$6 \tablenotemark{k}&\ssz 961215$-$961231&\ssz na&\ssz 05 53 02.97&\ssz $-$17 52 23.5\nl

\ssz 1997C&\ssz Ic&\ssz na&\ssz 970114&\ssz na&\ssz 961216$^\dag$$-$970104\tablenotemark{l}&\ssz na&\ssz 10 13 56.18&\ssz +38 49 00.5\nl

\ssz 1997X&\ssz Ic&\ssz .0037&\ssz 970201&\ssz 970125$\pm$6\tablenotemark{m \&\ssz  n}&\ssz 970114$^\dag$$-$970121&\ssz 961231$-$970121&\ssz 12 48 14.28&\ssz $-$03 19 58.5\nl

\ssz 1997dc&\ssz Ib&\ssz .0116&\ssz 970805&\ssz 970811$\pm$6 \tablenotemark{o}&\ssz 970719$-$970804&\ssz na&\ssz 23 28 28.41&\ssz +22 25 23.0\nl

\ssz 1997dq&\ssz Ib&\ssz .0033&\ssz 971102&\ssz 971105$\pm$6 \tablenotemark{p}&\ssz 971013$-$971029&\ssz na&\ssz 11 40 55.90&\ssz +11 28 45.7\nl

\ssz 1997ef&\ssz Ic&\ssz .0117&\ssz 971125&\ssz 971206$\pm$6 \tablenotemark{q \&\ssz  r}&\ssz 971116$-$971125*&\ssz 971113$-$971125&\ssz 07 57 02.87&\ssz +49 33 41.3\nl

\ssz 1997ei&\ssz Ic&\ssz .0106&\ssz 971223&\ssz 971225$\pm$10
\tablenotemark{s}&\ssz 971201$-$971223*&\ssz 971103$-$971203&\ssz 11 54 59.98&\ssz +58 29 26.4\tablebreak

\ssz 1998T&\ssz Ib&\ssz .0101&\ssz 980303&\ssz na&\ssz 980214$-$980303\tablenotemark{t}&\ssz 980208$-$980303&\ssz 11 28 33.16&\ssz +58 33 43.7\nl

\ssz 1998bo&\ssz Ic&\ssz .0161&\ssz 980422&\ssz 980412$\pm$10 \tablenotemark{u}&\ssz 980329$^\dag$$-$980412&\ssz na&\ssz 19 57 22.55&\ssz $-$55 08 18.4\nl

\ssz 1998cc&\ssz Ib&\ssz .0134&\ssz 980515&\ssz na &\ssz 980429$-$980515*\tablenotemark{v}&\ssz na&\ssz 13 29 19.31&\ssz +17 02 42.4\nl

\ssz 1998cv&\ssz Ic&\ssz .0272&\ssz 980624&\ssz 980627$\pm$6 \tablenotemark{w}&\ssz 980607$-$980623&\ssz na&\ssz 22 09 46.29&\ssz $-$49 47 43.0\nl

\enddata

\tablenotetext{\#}{z values computed from recession velocities found in
the Sternberg SN Catalog}

\tablenotetext{*}{Latest estimated explosion date limited by supernova
discovery date}

\tablenotetext{\dag} {Earliest estimated explosion date limited by
absence of supernova on image taken two days after this date}

\tablenotetext{a}{\cite{Williams1}\ \ 
$^{b}$\cite{Phillips1}\ \  
$^{c}$\cite{Iwamoto1}\ \ 
$^{d}$\cite{Benetti1}\ \ 
$^{e}$\cite{Filippenko1}\ \ 
$^{f}$\cite{Cappellaro1}\ \ 
$^{g}$\cite{Germany1}\ \ 
$^{h}$\cite{Benetti2}\ \ 
$^{i}$\cite{Pollas1}\ \ 
$^{j}$\cite{Filippenko2}\ \ 
$^{k}$\cite{Benetti3}\ \  
$^{l}$\cite{Li1}\ \ 
$^{m}$\cite{Nakano1}\ \
$^{n}$\cite{Benetti4}\ \ 
$^{o}$\cite{Piemonte1}\ \ 
$^{p}$\cite{Jha1}\ \ 
$^{q}$\cite{Iwamoto2}\ \
$^{r}$\cite{Garnavich1}\ \
$^{s}$\cite{Wang1}\ \ 
$^{t}$\cite{Li2}\ \ 
$^{u}$\cite{Patat1}\ \ 
$^{v}$\cite{Jha2}\ \
$^{w}$\cite{Phillips2}}

\end{deluxetable}

\begin{deluxetable}{lcllllll}
\tablewidth{500pt}

\tablecaption{
 \label{sngrb}
 Type Ib-Ic Supernovae and GRB Odds}
\tablehead{SN/GRB&Dates&RA&DEC&$\sigma^{total}$&$\Delta\theta$&Likelihood&Odds}

\startdata

SN 1992ar&920712$-$920727&23 17 28.40&$-$44 38 53.8& & & &0.5200\nl

GRB 920721& & 00 02 33 & $-$28 10 & 4.36 & 18.76 & $5.551\times
10^{-7}$ & \nl
\tablevspace{2mm}

SN 1993P&930501$-$930517&13 29 25.80&$-$30 24 47.4& & & &0.5200\nl

GRB 930510& & 10 02 40 &   +55 00 & 13.09 & 96.20 & $5.953\times
10^{-20}$ &\nl
\tablevspace{2mm}

SN 1994I&940326$-$940402&13 29 54.01&+47 11 31.7& & & &0.5236\nl

GRB 940331& & 10 31 14 &   +57 31 & 10.79 & 28.55 & $4.941\times
10^{-2}$ &\nl
\tablevspace{2mm}

SN 1994ai&941204$-$941220&02 23 06.17&$-$21 13 58.3& & & &0.5200\nl

GRB 941217& & 08 40 40 & $-$80 22 & 9.84 &  69.81 & $1.100\times
10^{-20}$ &\nl
\tablevspace{2mm}

SN 1995F&950114$-$950207&09 04 57.40&+59 55 58.7& & & &0.5200\nl

GRB 950118& & 06 51 52 & +58 17 & 4.36 &  16.97 & $2.470\times
10^{-5}$ &\nl
\tablevspace{2mm}

SN 1996D&960129$-$960209&04 34 00.00&$-$08 35 00.0& & & &0.5200\nl

GRB 960129& & 06 02 55 & $-$38 36 & 10.02 &  36.06 & $8.628\times
10^{-5}$ &\nl
\tablevspace{2mm}

SN 1996N&960214$-$960308&03 38 55.31&$-$26 20 04.1& & & &1.285\nl

GRB 960229& & 04 02 26 & $-$15 15 & 13.01 &  12.36 & $3.174\times
10^{1}$ &\nl
\tablevspace{2mm}

SN 1996aq&960731$-$960816&14 22 22.73&$-$00 23 24.3& & & &0.5210\nl

GRB 960731& & 13 18 07 & $-$18 17 & 8.52 &  23.86 & $2.908\times
10^{-2}$ &\nl
\tablevspace{2mm}

SN 1996cd&961129$-$961216&07 57 20.73&+11 12 23.7& & & &0.5200\nl

GRB 961216& & 04 38 50 & $-$21 28 & 6.48 &  58.66 & $1.498\times
10^{-35}$ &\nl
\tablevspace{2mm}

SN 1997B&961215$-$961231&05 53 02.97&$-$17 52 23.5& & & &1.901\nl

GRB 961218& & 06 31 00 & $-$21 43 & 15.31 &  9.72 & $4.052\times
10^{1}$ &\nl
\tablevspace{2mm}

\tablebreak

SN 1997C&961216$-$970104&10 13 56.18&+38 49 00.5& & & &0.5200\nl

GRB 961218& & 06 31 00 & $-$21 43 & 15.31 &  79.90 & $1.981\times
10^{-10}$ &\nl
\tablevspace{2mm}

SN 1997X&970114$-$970121&12 48 14.28&$-$03 19 58.5& & & &0.5200\nl

GRB 970116& & 08 13 48 & $-$11 09 & 6.66 &  68.38 & $7.724\times
10^{-45}$ &\nl
\tablevspace{2mm}

SN 1997dc&970719$-$970804&23 28 28.41&+22 25 23.0& & & &0.5200\nl

GRB 970725& & 00 51 47 & +36 12 & 6.04 &  22.72 & $4.517\times
10^{-5}$ &\nl
\tablevspace{2mm}

SN 1997dq&971013$-$971029&11 40 55.90&+11 28 45.7& & & &1.718\nl

GRB 971013& & 11 08 07 & +02 39 & 11.62 &  12.00 & $3.299\times
10^{1}$ &\nl
\tablevspace{2mm}

SN 1997ef&971116$-$971125&07 57 02.87&+49 33 41.3& & & &0.5288\nl

GRB 971120& & 10 23 02 & +76 24 & 12.71 & 30.42 & $1.517\times
10^{-1}$ &\nl
\tablevspace{2mm}

SN 1997ei&971201$-$971223&11 54 59.98&+58 29 26.4& & & &0.8908\nl

GRB 971220& & 13 41 45 & +58 50 & 9.44 &  13.79 & $1.475\times
10^{1}$ &\nl
\tablevspace{2mm}

SN 1998T&980214$-$980303&11 28 33.16&+58 33 43.7& & & &0.5200\nl

GRB 980223& & 14 39 23 & +27 56 & 10.44 &  44.78 & $2.706\times
10^{-7}$ &\nl
\tablevspace{2mm}

SN 1998bo&980329$-$980412&19 57 22.55&$-$55 08 18.4& & & &0.5200\nl

GRB 980404& & 04 36 57 & $-$49 57 & 12.22 &  66.91 & $4.882\times
10^{-12}$ &\nl
\tablevspace{2mm}

SN 1998cc&980429$-$980515&13 29 19.31&+17 02 42.4& & & &0.8349\nl

GRB 980501& & 14 55 28 & +23 21 & 17.33 &  21.15 & $9.258\times
10^{0}$ &\nl
\tablevspace{2mm}

SN 1998cv&980607$-$980623&22 09 46.29&$-$49 47 43.0& & & &0.5200\nl

GRB 980609& & 21 22 04 & $-$18 38 & 3.83 &  32.58 & $7.241\times
10^{-33}$ &\nl

\enddata

\end{deluxetable}

\begin{deluxetable}{lll}

\tablecaption{\label{results} Summary Of Results of Bayesian Analysis}

\tablehead{\colhead{SN Sample} & \colhead{Model} & \colhead{Odds}}

\startdata

All Ia & $f=1$ & $2.23\times 10^{-22}$ \nl
Ia ($z\le 0.1$) & $f=1$ & $3.01\times 10^{-10}$ \nl
Ia ($z>0.1$) & $f=1$ & $7.41\times 10^{-13}$ \nl
All Ia & $f\le 1$ & $2.90\times 10^{-2}$ \nl
Ia ($z\le 0.1$) & $f\le 1$ & $6.24\times 10^{-2}$\nl
Ia ($z>0.1$) & $f\le 1$ & $5.22\times 10^{-2}$\nl
Ib, Ib/c, Ic & $f=1$ & $1.76\times 10^{-4}$ \nl
Ib, Ib/c, Ic (W\&W)  & $f=1$ & $1.62\times 10^{-1}$ \nl
Ib, Ib/c, Ic & $f\le 1$ & $1.59\times 10^{-1}$ \nl

\enddata

\end{deluxetable}


\clearpage

\begin{thebibliography}{}

\bibitem[Band \& Hartmann 1998]{bh98}
Band,~D.~L., and Hartmann,~D.~H. 1998, ApJ, 493, 555


\bibitem[Benetti 1994]{Benetti1} 
Benetti, S. 1994, IAU Circ. No. 6120

\bibitem[Benetti, Turatto, \& Augusteijn 1996]{Benetti2} 
Benetti, S., Turatto, M., \& Augusteijn, T. 1996, IAU Circ. No. 6454

\bibitem[Benetti \& Lidman 1997]{Benetti3} 
Benetti, S. \& Lidman, C. 1997, IAU Circ. No. 6535

\bibitem[Benetti, Turrato, \& Perez 1997]{Benetti4} 
Benetti, S., Turrato, M., \& Perez, I. 1997, IAU Circ. No. 6554

\bibitem[Cappellaro \& Pata 1996]{Cappellaro1} 
Cappellaro, E., \& Pata, F. 1996, IAU Circ. No. 6317 

\bibitem[CfA SN Team Website 1998]{CfA1} 
CfA Supernova Search Team Website 1998,
http://cfa-www.harvard.edu/cfa/oir/Research/supernova/RecentSN.html

\bibitem[Costa et al. 1997a]{costa97a}
Costa, E., Feroci, M., Frontera, F., Zavattini, G., Nicastro, L.,
Palazzi, E., Spoliti, G., Di Ciolo, L., Coletta, A., D'Andreta, G et
al. 1997a, IAU Circ 6572

\bibitem[Costa et al. 1997b]{costa97b}
Costa, E., Feroci, M., Piro, L., Cinti, M.~N., Frontera, F., Zavattini,
G., Nicastro, L., Palazzi, E., Dal Fiume, D., Orlandini, M. et al.
1997b, IAU Circ 6576

\bibitem[Della Valle 1993]{Della Valle1} 
Della Valle, M. 1993, IAU Circ. No. 5809

\bibitem[Djorgovski et al. 1998]{djorgovski98}
Djorgovski, S.~G., Kulkarni, S.~R., Goodrich, R., Frail, D.~A. \&
Bloom, J.~S. et al. 1998, GCN Report 139

\bibitem[Filippenko 1992]{filippenko92} 
Filippenko, A.V. 1992, IAU Circ. No. 5555

\bibitem[Filippenko \& Barth 1995]{Filippenko1} 
Filippenko, A. V., \& Barth, A.J. 1995, IAU Circ. No. 6138

\bibitem[Filippenko, et al. 1997]{Filippenko2} 
Filippenko, A. V., Leonard, D. C., Gilbert, A. M., \& Ho, W. C. G. 
1997, IAU Circ. No. 6577

\bibitem[Galama et al. 1998]{galama98}
Galama, T.~J. et al. 1998, Nature, submitted (astro-ph/9806175)

\bibitem[Garnavich, et al. 1997]{Garnavich1} 
Garnavich, P., Jha, S., Kirshner, R., Challis, P., Balam, D., 
Brown, W., \& Briceno, C. 1997 IAU Circ. No. 6786

\bibitem[Garnavich, Challis, \& Kirshner 1996]{Garnavich2} 
Garnavich, P., Challis, P., \& Kirshner, R. 1996, IAUC No. 6424

\bibitem[Germany, et al. 1996]{Germany1} 
Germany, L., Schmidt, B., Stathakis, R., \& Johnston, H. 1996, 
IAU Circ. No. 6351

\bibitem[Graziani et al. 1992]{graziani92} 
Graziani,~C., Lamb,~D.~Q., Loredo,~T.~J., Fenimore,~E.~E.,
Murakami,~T., and Yoshida,~A. 1992, in Compton Gamma-Ray Observatory,
ed. M.~Friedlander, N.~Gehrels, and D.~Macomb (New York: AIP) 897

\bibitem[Graziani \& Lamb 1996]{graziani96}  
Graziani,~C. \& Lamb,~D.~Q. 1996, in Gamma-Ray Bursts, AIP Conf.
Proceedings No. 384, ed. C. Kouveliotou, M. F. Briggs \& G. J. Fishman
(New York: AIP), 382

\bibitem[Groot et al. 1997]{groot97}
Groot. P., Galama, T., van Paradijs, J., Strom, R., Telting, J.,
Rutten, R.~G.~M., Pettini, M.,Ttanvir, N. et al. 1997, IAU Circ 6584

\bibitem[Hakkila et al. 1998]{hakkila98} Hakkila,~J., Meegan,~C.~A.,
Pendleton,~G.~N., Henze,~W., McCollough,~M., Kommers,~J.~M., and
Briggs,~M.~S. 1998, in Gamma-Ray Bursts, AIP Conference Proceedings
428, eds. C. A. Meegan, R. D. Preece, \& T. M. Kogut (New York: AIP),
509

\bibitem[H\"oflich, Wheeler \& Wang 1998]{hoflich98}
H\"oflich, P., Wheeler, J.~C. \& Wang, L. 1998, ApJ, submitted 
(astro-ph/9807345)

\bibitem[Hurley et al. 1998]{hurley98}
Hurley, K. et al. 1998, ApJS, submitted (data are publicly available at 
http://ssl.berkeleyedu/ipn3.index.html

\bibitem[Iwamoto, et al. 1994]{Iwamoto1} 
Iwamoto, K., Nomoto, K., Hoeflich, P., Yamaoka, H., Kumagai, S., 
\& Shigeyama, T. 1994, ApJ, 437, L114

\bibitem[Iwamoto et al. 1998]{iwamoto98}
Iwamoto, K. et al. 1998, Nature, submitted (astro-ph/9806322)

\bibitem[Iwamoto, et al. 1998]{Iwamoto2} 
Iwamoto, K., Nakamura, T., Nomoto, K., Mazzali, P., Garnavich, P., 
Kirshner, R., Jha, S., \& Balam, D. 1998, ApJ, submitted
(astro-ph/9807060)

\bibitem[Jha, et al. 1997]{Jha1} 
Jha, S., Challis, P., Garnavich, P., \& Kirshner, R. 1997, 
IAU Circ. No. 6770

\bibitem[Jha, Garnavich, \& Kirshner 1998]{Jha2} 
Jha, S., Garnavich, P., \& Kirshner, R. 1998, IAU Circ. No. 6907

\bibitem[Kippen et al. 1998a]{kippen98a}
Kippen, R. M. and the BATSE team 1998, GCN Report 67

\bibitem[Kippen et al. 1998b]{kippen98b} 
Kippen, R.~M., Briggs, M.~S., Kommers, J.~M., Kouveliotou, C., Hurley,
K., Robinson, C.~R., van Paradijs, J., Hartmann, D.~H., Galama, T.~J.,
\& Vreeswijk, P.~M. 1998, ApJ, submitted (astro-ph/9806364)

\bibitem[Kirshner, et al. 1995]{Kirshner1} 
Kirshner, R., Jayawardhana, R., Filippenko, A. V., Perlmutter, S.,
Barth, A. J., \& Hook, I 1995, IAU Circ. No. 6267

\bibitem[Kulkarni et al. 1998a]{kulkarni98a}
Kulkarni, S.~R. et al. 1998, Nature, 393, 35

\bibitem[Kulkarni et al. 1998b]{kulkarni98b}
Kulkarni, S.~R., Frail, D.~A., Wieringa, M.~H., Ekers, R.~D., Sadler,
E.~M., Wark, R.~M., Higdon, J.~L., Phinney, E.~S. \& Bloom, J.~S. 1998,
Nature, submitted (astro-ph/9807001)

\bibitem[Li, et al. 1997]{Li1} 
Li, W., Qiao, Q., Qiu, Y., \& Hu, J. 1997, IAU Circ. No. 6536

\bibitem[Li, Li, \& Wan  1998]{Li2} 
Li, W., Li, C., \& Wan, Z. 1998, IAU Circ. No. 6830

\bibitem[Loredo \& Lamb 1992]{ll92} 
Loredo,~T.~J., and Lamb,~D.~Q. 1992, in Gamma-Ray Bursts, ed.
W.~Paciesas and G.~J.~Fishman (Woodbury, N.Y.: AIP), 414

\bibitem[Mardia 1972]{mardia} 
Mardia,~K.~V. 1972, Statistics of Directional Data (London: Academic
Press)

\bibitem[McNaught 1992]{mcnaught92}
McNaught, R.~H. 1992, IAU Circ. No. 5552

\bibitem[Meegan et al. 1992]{meegan92}
Meegan, C. A. et al. 1992, Nature, 355, 143

\bibitem[Meegan et al. 1998]{meegan98}
Meegan, C. A. et al. 1998, in Gamma-Ray Bursts, AIP Conference Proceedings
428, eds. C. A. Meegan, R. D. Preece, \& T. M. Kogut (New York: AIP),
3

\bibitem[Metzger et al. 1997]{metzger97}
Metzger, M.~R., Djorgovski, S.~G., Kulkarni, S.~R., Steidel, C.~C.,
Adelberger, K.~L., Frail, D.~A., Costa, E., \& Frontera, F. 1997c,
Nature, 387, 878

\bibitem[Nakano \& Aoki 1997]{Nakano1} 
Nakano, S. \& Aoki, M. 1997, IAU Circ. No. 6552

\bibitem[Norris, Bonnell \& Watanabe 1998]{norris98}
Norris, J.~P., Bonnell, J.~T. \& Watanabe, K. 1998, ApJ, submitted 
(astro-ph/9807322)

\bibitem[Paczy\'nski 1998]{paczynski98}
Paczy\'nski, B. 1998, ApJ, 494, L45

\bibitem[Patat \& Maia 1998]{Patat1} 
Patat, F. \& Maia, A. 1998, IAU Circ. No. 6889

\bibitem[Perlmutter, et al. 1998]{Perlmutter1}  
Perlmutter, S., et al. Ap.J., (submitted)

\bibitem[Phillips 1993a]{Phillips3} 
Phillips, M. 1993a, IAU Circ. No. 5699

\bibitem[Phillips 1993b]{Phillips1} 
Phillips, M. 1993b, IAU Circ. No. 5799

\bibitem[Phillips 1998]{Phillips2} 
Phillips, M. 1998, IAU Circ. No. 6968

\bibitem[Pian et al. 1998a]{pian98a}
Pian, E., Antonelli, L.~A., Daniele, M.~R., Rebecchi, S., Torroni, V.,
Gennaro, G, Ferorci, M. \& Piro, L. 1998, GCN Report 61

\bibitem[Pian et al. 1998b]{pian98b}
Pian, E., Frontera, F., Antonelli, L.~A. \& Piro, L. 1998, GCN Report 
69

\bibitem[Piemonte, Benetti, \& Turatto 1997]{Piemonte1} 
Piemonte, A., Benetti, S., \& Turatto, M. 1997, IAU Circ. No. 6717

\bibitem[Piro, et al. 1998]{piro98}
Piro,~L., Butler,~R.~C., Fiore,~F., Antonelli,~A., \& Pian,~E. 1998, GCN
Report 155

\bibitem[Pollas 1997]{Pollas1} 
Pollas, C. 1997, IAU Circ. No. 6531

\bibitem[Riess, et al. 1998]{Riess1} 
Riess, A., Nugent, P., Filippenko, A., Kirshner, R., \& Permlmutter, S.
1998, ApJ, in press (astro-ph/9804065)

\bibitem[Smith \& Lamb 1993]{smith93}
Smith, I. A. \& Lamb, D. Q. 1993, ApJ, 410, L23

\bibitem[Soffitta, P. et al. 1998]{soffitta98}
Soffitta, P. et al. 1998, IAU Circ. No. 6884

\bibitem[Strom 1995]{strom95}
Strom, R~G. 1995, in The Lives of the Neutron Stars, ed. M.~A. Alpar,
\"U. Kiziloglu and J. van Paradijs (NATO ASI Series: Kluwer) Series C,
450,

\bibitem[Tinney et al. 1998]{tinney98}
Tinney, C., Stathakis, R., Cannon, R., Galama, T.~J. 1998, 
IAU Circ. No. 6896

\bibitem[Wang, Howell, \& Wheeler 1998]{Wang1} 
Wang, L., Howell, D. A., \& Wheeler, J. C. 1998, IAU Circ. No. 6802

\bibitem[Wang \& Wheeler 1998]{wang98}
Wang, L. \& Wheeler, J.~C.  1998, ApJ, in press (astro-ph/9806212)

\bibitem[Waxman \& Loeb 1998]{waxman98}
Waxman, E. \& Loeb, A. 1998, ApJ, submitted (astro-ph/9808135)

\bibitem[Williams, Hamuy, \& Phillips 1992]{Williams1} 
Williams, R., Hamuy, M., \& Phillips, M. 1992, IAU Circ. No. 5574

\bibitem[Woosley, Eastman \& Schmidt 1998]{woosley98}
Woosley, S.~E., Eastman, R.~G. \& Schmidt, B.~P. 1998, ApJ, submitted
(astro-ph/9806299)

\bibitem[Woosley \& Weaver 1986]{woosley86}
Woosley, S.~E. \& Weaver, T.~A. 1986, ARA\&A, 24, 205

\end{thebibliography}
\end{document}